\documentclass[useAMS,usenatbib]{mn2e}
\usepackage{graphics}
\usepackage{epsfig}

\newcommand{\kepler}[1]{\textit{Kepler}}

\title[Kepler Observations of Sco X-1]{Kepler K2 Observations of Sco X-1: Orbital
Modulations and Correlations with Fermi GBM and MAXI}

\author[R. I. Hynes et al.]{Robert~I. Hynes$^1$, Bradley~E. Schaefer$^1$,
  Zachary~A. Baum$^1$, Ching-Cheng Hsu$^1$,\newauthor Michael~L. Cherry$^1$,
  Simone Scaringi$^2$\\
$^{1}$Department of Physics and Astronomy, Louisiana State
University, Baton Rouge, Louisiana 70803, USA\\ 
$^2$Max Planck Institute für Extraterrestriche Physik, D-85748 Garching, Germany}
\begin{document}

\date{Submitted}

\pagerange{\pageref{firstpage}--\pageref{lastpage}} \pubyear{2015}

\maketitle

\label{firstpage}

\begin{abstract}

  We present a multi-wavelength study of the low-mass X-ray binary
  Sco~X-1 using {\it Kepler} K2 optical data and {\it Fermi} GBM and
  MAXI X-ray data. We recover a clear sinusoidal orbital modulation
  from the {\it Kepler} data. Optical fluxes are distributed bimodally
  around the mean orbital light curve, with both high and low states
  showing the same modulation. The high state is broadly consistent
  with the flaring branch of the Z diagram and the low state with the
  normal branch. We see both rapid optical flares and slower dips in
  the high state, and slow brightenings in the low state. High state
  flares exhibit a narrow range of amplitudes with a striking cut-off
  at a maximum amplitude.  Optical fluxes correlate with X-ray fluxes
  in the high state, but in the low state they are
  anti-correlated. These patterns can be seen clearly in both flux-flux
  diagrams and cross-correlation functions and are consistent between
  MAXI and GBM. The high state correlation arises promptly with at
  most a few minutes lag. We attribute this to thermal reprocessing of
  X-ray flares. The low state anti-correlation is broader, consistent
  with optical lags of between zero and 30~min, and strongest with
  respect to high energy X-rays. We suggest that the decreases in
  optical flux in the low state may reflect decreasing efficiency of
  disc irradiation, caused by changes in the illumination
  geometry. These changes could reflect the vertical extent or
  covering factor of obscuration or the optical depth of scattering
  material.
\end{abstract}

\begin{keywords}
accretion, accretion discs - X-rays: binaries - X-rays: individual: Sco
X-1
\end{keywords}

\section{Introduction}
\label{IntroSection}

Scorpius~X-1 (Sco~X-1) was the first extra-Solar X-ray source to be
discovered \citep{Giacconi:1962a} and remains a challenging object to
explain after over fifty years of study.  It is a low-mass X-ray
binary (LMXB) and the prototype of the subclass of Z sources, so-named
after the locus of points in an X-ray colour-colour diagram
\citep{Hasinger:1989a}.  Z sources are generally believed to be
accreting at high rates, near or above the Eddington limit, as
compared to the more numerous atoll sources which accrete at lower
rates.  They are subdivided into Sco-like systems, behaving like
Sco~X-1 and Cyg-like systems similar to Cyg~X-2.  Various attempts
have been made to associate these different classes and subclasses of
neutron star LMXB with other distinctive characteristics, for example
the neutron star magnetic field or our viewing angle, but these are
difficult to reconcile with the behavior of the transient LMXB
XTE~J1701--462 which was observed to move from Cyg-like Z source to
Sco-like Z source to atoll source during the decay of its outburst
\citep*{Lin:2009a,Homan:2010a}.  This suggests that the primary
distinguishing feature between the atoll class and the two Z
subclasses is simply the accretion rate.

The three branches of the Z diagram are the horizontal branch (HB),
the normal branch (NB), and the flaring branch (FB), with Sco~X-1
spending most of its time on the NB and FB.  The HB-NB transition is
referred to as the hard apex and the NB-FB transition is the soft
apex. The Z diagram is often assumed to be an monotonically increasing
sequence in mass accretion rate from HB to FB
\citep*[e.g.][]{Psaltis:1995a}, but this is challenged by some models
with \citet{Church:2012a} arguing that the lowest mass transfer rate
occurs at the soft apex, with both the NB and the FB representing
different modes of increasing accretion, while \citet*{Lin:2010a}
infer no change in mass transfer rate at all around the diagram. There
is currently, no widely accepted consensus on the correct
interpretation of the Z diagram.

\begin{figure}
\includegraphics[width=3.2in,trim=0.0in 0.0in 2.5in 7.2in, clip]{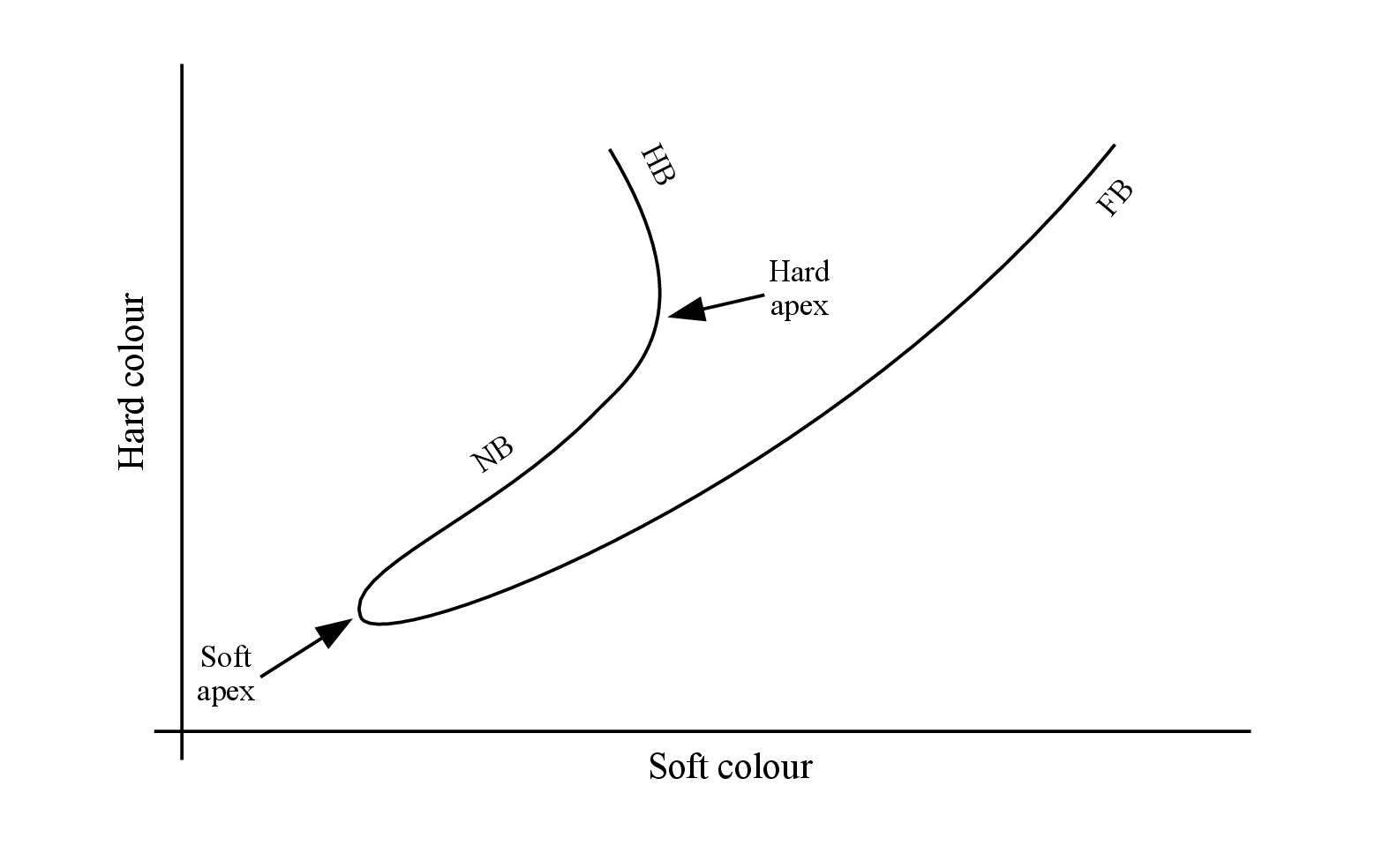}
\caption{Schematic Z diagram for Sco~X-1 based on
  \citet{Hasinger:1989a}. The three branches are annotated HB
  (horizontal branch), NB (normal branch) and FB (flaring
  branch). Note that the HB is short, and nearly vertical in Sco-like
  Z sources.}
\label{SchematicFig}
\end{figure}

The optical counterpart to Sco~X-1, V818~Sco, was discovered by
\citet{Sandage:1966a}, opening the door for subsequent multi-wavelength
studies aiming to study optical variability and relate X-ray and
optical behavior
\citep{Hiltner:1967a,Hiltner:1970a,Mook:1975a,Bradt:1975a,Canizares:1975a}.
This early work established large amplitude optical variability, with
several preferred optical flux levels resulting in bimodal or trimodal
flux histograms.  At fainter optical levels, optical variations
appeared uncorrelated with X-rays, or at times anti-correlated, but
when the source was brightest a correlation emerged with the the large
X-ray variations in the FB.  Subsequent similar studies
\citep[e.g.][]{Augusteijn:1992a,McNamara:2003a} have confirmed these
patterns with better data.

The discovery of optical and X-ray correlations at some times also
raises the possibility of using simultaneous rapid data to measure
lags between the X-ray and optical, on the assumption that optical
variability is produced by thermal reprocessing of incident X-rays by
the disc and companion star.  Lags of order seconds are expected due
to light travel times within the binary and can be used to identify
where light is being reprocessed and possibly even measure system
parameters \citep{OBrien:2002a}.  The first thorough studies were by
\citet{Ilovaisky:1980a} and \citet{Petro:1981a}.  Both groups found
that the optical was correlated, and lagged behind the X-rays with
some smearing of the reprocessed variability.  \citet{Petro:1981a}
found that time-scales less than about 20~s were smoothed out.
\citet{McGowan:2003a} performed a more sophisticated analysis of these
data sets, but found that in some cases the optical could not be
straightforwardly described with a simple reprocessing model.  Where
this method did work, they found a lag of $8.0\pm0.8$~s and smearing
of $8.6\pm1.3$~s.  \citet{MunozDarias:2007a} performed a modern study
using fast CCDs and narrow-band filters simultaneously with the {\it
  Rossi X-ray Timing Explorer}.  They found lags of 11--16~s
associated with the Bowen blend emission lines around 4640~\AA\ at the
top of the FB, and attributed these to reprocessing on the companion
star.  Correlations were also found at the bottom of the FB with lags
of 5--10~s, which were attributed to reprocessing in the disc.
\citet{Britt:2013a} performed an extensive study spanning all orbital
phases and a range of states, and found that correlations were only
present when Sco~X-1 was in the FB.  No orbital modulation was seen in
the lags, suggesting the reprocessed signal was dominated by the disc,
at least in the continuum.

The binary period was identified as 18.9~hr based on the discovery of
a photometric modulation by \citet*{Gottlieb:1975a} in archival
photographic plates and spectroscopically confirmed by
\citet{Cowley:1975a}.  Because of the large amplitude aperiodic
variability in the source, the orbital modulation cannot be seen in
individual light curves and only emerges when large data sets are
folded \citep{Gottlieb:1975a,Augusteijn:1992a,Hynes:2012a}.  The
photometric modulation arises from X-ray heating of the inner face of
the donor star and takes the form of a sinusoidal modulation of full
amplitude 0.13--0.26~mag.  \citet{Augusteijn:1992a} found that the
amplitude and phasing of modulation was the same in both high and low
states, but no data set has been sufficiently comprehensive to permit
detailed study of the light curve morphology.  Irradiation of the
donor star also produces narrow emission lines of N\,{\sc iii} and
C\,{\sc iii} (the 4640~\AA\ Bowen blend noted above) phased as
expected based on the photometric modulation
\citep{Steeghs:2002a}. These provide the potential to determine the
system parameters, although since they do not trace the motion of the
center of mass, and we remain uncertain about the binary mass ratio
and inclination, there are substantial uncertainties.  The most
comprehensive analysis has been performed by \citet{MataSanchez:2015a}
who conclude that the neutron star must have a mass less than
1.73~M$_{\odot}$ at 90~percent confidence.

A key ingredient in interpreting observations of any LMXB is the
binary inclination.  In Sco~X-1, this is generally taken to be low.
\citet*{Fomalont:2001a} estimated $i=44^{\circ}\pm6^{\circ}$ from
modelling of the orientation of the twin radio jets.  This method does
not in general give a reliable measure of the orbital inclination as
the jets may be misaligned with the accretion disc.  For example in
two microquasars there is a clear discrepancy between inclinations
derived from jets and those from optical data \citep{Maccarone:2002a}
and in SS433 we observe precession of the radio jets on a 164~d period
attributed to misalignment between the jet and the orbital axis
\citep{Hjellming:1981a}.  On the other hand in Sco~X-1
\citet{Fomalont:2001a} find the jet inclination to be stable over five
years, implying negligible precession and a jet aligned with the
orbital plane, suggesting that this measurement may be reliable.
\citet{MataSanchez:2015a} independently examine system parameters
derived from optical spectroscopy and conclude that an inclination
higher than $40^{\circ}$ would require a neutron star of mass below
1.4~M$_{\odot}$.

Beginning in 2009 May, the {\it Kepler} mission monitored a fixed area
of the sky to search for extra-Solar planets \citep{Koch:2010a}.  With
the failure of a second reaction wheel in 2013 May, the primary
mission came to an end as precise pointing could no longer be
maintained.  {\it Kepler} received a second wind in the form of the K2
mission \citep{Howell:2014a}.  Pointing could be maintained with the
aid of the Solar wind provided the satellite only pointed at targets
on the ecliptic, and so the K2 mission has taken the form of a series
of three month campaigns spaced around the ecliptic.  Campaign 2 was
observed from 2014 August 23 to November 13, and fortuitously included
Sco~X-1 in the field.  We report here our analysis of this data set in
conjunction with simultaneous X-ray monitoring with the {\it Fermi}
GBM and MAXI.  \citet{Scaringi:2015a} and \citet{Hakala:2015a} report
independent analyses using {\it Kepler} and MAXI only.  In
Section~\ref{ObservationSection} we describe the data sets used.  We
analyse the orbital modulation in Section~\ref{OrbitalSection},
allowing us to remove its effects and examine the residual variability
in Section~\ref{LightcurveSection}.  Section~\ref{FlareSection}
examines the statistical properties of individual flares identified
when the source is brightest. We proceed to incorporate X-ray data
into the analysis in the form of X-ray/optical flux-flux diagrams in
Section~\ref{FluxDiagramSection} and then examine lags in
Section~\ref{CCFSection}. Finally in Section~\ref{XraySpecSection} we
use states identified in {\it Kepler} data to sort GBM and MAXI data
by spectral state and hence construct average spectra for each
state. We round up by discussing some interpretations of our results
in Section~\ref{DiscussionSection}, propose a possible mechanism for
relating the X-ray and optical behavior in Section~\ref{ModelSection},
and finally summarize our findings and conclusions in
Section~\ref{ConclusionSection}.

\section{Observations}
\label{ObservationSection}

\subsection{Kepler}
\label{KeplerSection}

{\it Kepler} observed Sco~X-1 in short-cadence mode, with 54.2~s
integrations, from Julian dates (JD) 2\,456\,893.275 (2014 August 23)
to 2\,456\,972.058 (2014 November 10), for a total of close to 78.8~d,
with few gaps in a relentless cadence.  These {\it Kepler}
observations, as part of the K2 mission, were proposed, accepted, and
observed for us under K2 proposal GO2026 (R. Hynes principal
investigator) and independently by other teams.  The images were sent
to the ground with a small `postage stamp' of 14 by 12~pixels (with
7~pixels in the corners missing).  Each pixel is 4~arcsec square, with
the detector resolution being somewhat smaller, so the star images are
under-sampled.  Sco~X-1 is the only star in the field.  The entire
{\it Kepler} experience is that the stability of the `absolute'
photometry is much better than the parts-per thousand level.  The
photometric aperture chosen was a 41~pixel area that covered
effectively all the Sco~X-1 light, with this aperture being
sufficiently large so that the usual small movements of the {\it
  Kepler} spacecraft made for no flux shifting in-and-out of the
aperture. Since {\it Kepler} data are flat-fielded with a pre-launch
flat, there potentially are variations introduced by motion of the
source with respect to secular changes in the flat field. These are
seen in other K2 targets, but are below 1~percent and negligible
compared to the intrinsic source variability. The typical source flux
is around 140\,000~e$^{-}$~s$^{-1}$, so the Poisson one-sigma
uncertainty is around the 0.04~percent level per integration. This
photometric uncertainty is also negligibly small compared to the
unresolved rapid variations of Sco~X-1.  Except where artefacts noted
below occur, we then expect {\em all} the variability seen to be
intrinsic to Sco~X-1 itself. Further evidence for this is the absence
of a white noise component even at the highest frequencies in the
power spectra shown by \citet{Scaringi:2015a}.

The flux of Sco~X-1 was extracted from the images with the usual {\it
  Kepler} pipeline.  The output has the flux inside our 41-pixel
photometry aperture, with a background subtraction from the rest of
the image, all reported in e$^{-}$~s$^{-1}$.  The wavelength
sensitivity is that of the Kepler CCD with no filters, resulting in a
so-called `{\it Kepler} magnitude' that covers a broad band over the
usual $B$, $V$ and $R$ range.  There is no easy or accurate means to
convert from the {\it Kepler} magnitude to any standard magnitude for
an object such as Sco~X-1 with a non-stellar spectral energy
distribution. Some variation in apparent brightness might be
introduced by changes in spectral shape for such a broad
filter. \citet{Augusteijn:1992a} found a difference of $0.024\pm0.005$
in the $(B-V)$ color of Sco~X-1 between optically bright and faint
states. Since this corresponds to only about a 2~percent variation in
$F_V / F_B$, we expect color changes to make a negligible contribution
to the variations we observe.

The {\it Kepler} light curves are generated with time stamps for the
middle of the exposure expressed as a barycentric Julian date (BJD) in
the barycentric dynamical time (TDB) system. Hereafter we will follow
\citet*{Eastman:2010a} and refer to such dates in the format
BJD$_{\rm TDB}$. Throughout this work, we will use this notation to
indicate both the date convention and time system used for original
data products and those we actually combine and analyze. Neglecting
the time system results in errors of order a minute which is not
negligible compared to the {\it Kepler} exposure times. A thorough
discussion of the importance of specifying time systems and
definitions of the systems used in this paper is provided by
\citet{Eastman:2010a}.

We know of three data artefacts that can affect our light curve.
First, images can have discrepant fluxes during occasional reaction
wheel angular momentum desaturation events due to the movement of the
Sco~X-1 image.  Fortunately, all such instances are flagged in the
{\it Kepler} pipeline output.  All fluxes taken during times of
desaturation events have been deleted from our data set together with
periods of coarse pointing around them.  In addition, time bins
following a desaturation event occasionally display clearly discrepant
fluxes, even when not flagged, and these have also been systematically
deleted.  In all, we discarded 1956 data points.  Second, in the usual
way, cosmic rays occasionally hit within a photometric aperture,
leaving apparent extra flux.  In other {\it Kepler} data sets, we see
that this happens in a given aperture on the order of a few times per
day, with detected events ranging from 20 to 400 counts.  This is
completely negligible in our Sco X-1 light curve.  Third, during the
time period BJD$_{\rm TDB}$ 2\,456\,911.23 to 2\,456\,912.05, solar
flares made for some extra flux in the photometry aperture.  Our
experience with Kepler point sources with no fast variability shows
that the effects are random enhancements of up to 1000 counts for some
time bins during this interval.  Such excess is negligibly small
compared to the fast variability in the base flux from Sco~X-1, so we
have not excised any light curve points during this time interval of
Solar activity.  In all, we have 115\,680 fluxes, all tied to an
accurate BJD$_{\rm TDB}$, to form our Kepler light curve, with
typically better than one percent photometric accuracy.

The {\it Kepler} light curve creates some difficulties in estimating
uncertainties.  Because of the exquisite data quality, instrumental
errors are negligible compared to the intrinsic variability in the
source.  This means that the noise characteristics are those of the
source, not of white noise.  In places, the light curve is dominated
by short flares of increasing flux (only), so the local flux
distribution is asymmetric and distinctly non-Gaussian.  In other
places, there is little very short time-scale variability.
Furthermore, the intrinsic time-scale of the variability introduces an
auto-correlation into the data.  The result is that the uncertainties
on the data points are not independently and identically distributed,
and are non-Gaussian.  This invalidates many usual statistical methods
(e.g.\ $\chi^2$ fitting).  A more robust approach that uses the
distribution within the data set itself is the bootstrap
\citep{Efron:1979a}, where data points are randomly drawn from the
light curve, with replacement, to create many resampled versions of
the light curve, which are then analyzed identically to the data.  In
this case, we require a bivariate bootstrap where we resample (time,
rate) pairs, as we need to preserve the timing information if we are
to analyze periodic modulations or correlations with other wavebands.

The simple bootstrap method is known to significantly underestimate
uncertainties when the data are autocorrelated.  In this case, the
most commonly used modification is the block bootstrap
\citep{Kunsch:1989a,Liu:1992a}.  In this case, rather than resampling
individual points, we resample blocks of consecutive points with the
length of block chosen to preserve the autocorrelation characteristics
of the data.  There are a variety of specific implementations of this
idea. For this work we adopt the stationary bootstrap of
\citet{Politis:1994a} which uses a geometric distribution of block
lengths and wraps around from the end of the time-series back to the
beginning. After testing this method on simulated data sets with the
same sampling, sinusoidal modulation, and noise power-spectrum to
Sco~X-1, we adopted an average block length of 2880 points,
corresponding to about 2~d. This is longer than the orbital period and
the typical time-scales of variability and in simulations recovers
correct uncertainties.

\subsection{Fermi GBM}

The Gamma-ray Burst Monitor (GBM) is one of the two instruments on
board the {\it Fermi Gamma-Ray Space Telescope}. It consists of 14
detectors: 12 NaI detectors and 2 BGO detectors. Typically 3--4 NaI
detectors view an Earth occultation within 60$^{\circ}$ of the
detector normal vector. The two BGO detectors are located on opposite
sides of the spacecraft and view a large part of the sky in the energy
range 150~keV to 40~MeV. None of the GBM detectors has direct imaging
capability. GBM has two continuous data types: CTIME data with nominal
0.256~s time resolution and 8-channel spectral resolution, and CSPEC
data with nominal 4.096~s time resolution and 128-channel spectral
resolution. The results presented in this paper use the lower spectral
resolution CTIME data and the higher spectral resolution CSPEC data
from the NaI detectors for the light curves and spectral studies
respectively.

{\it Fermi} was launched in June 2008 into a 25.6$^{\circ}$
inclination orbit at an altitude of 555~km. The diameter of the Earth
as seen from {\it Fermi} is 135$^{\circ}$, so roughly 30~percent of
the sky is occulted by the Earth at any one time. One complete orbit
of the spacecraft allows over 85~percent of the sky to be
observed. The Earth Occultation Technique \citep{WilsonHodge:2012a}
allows GBM to monitor the fluxes of known point sources. The technique
involves fitting a model consisting of a quadratic background plus
source terms to a short 4~min window of data centered on the
occultation time of the source of interest. The precession of the
orbital plane allows the entire sky to be occulted every 26~d (half
the precession period for the {\it Fermi} orbit), though the exposure
is not uniform.  All GBM data for Sco~X-1 coinciding with the {\it
  Kepler} observations were analyzed. In order to maintain data
quality, we apply several event selection cuts, for instance
eliminating occultations with high satellite spin rate and data from
South Atlantic Anomaly crossings.  Other cuts were applied to remove
occultations that coincide with solar flares.  The times of the solar
flares are determined by the {\it GOES} satellite.

The original occultation light curve had time stamps for the middle of
the occultation in mission time.  We have corrected this to
terrestrial time (TT) heliocentric Julian dates (HJD$_{\rm TT}$),
which are within four seconds of the BJD$_{\rm TDB}$ used for the {\it
  Kepler} light curve \citep{Eastman:2010a}.  Our observations extend
over a 78.8~d interval, so a correction to the heliocenter or the
barycenter is required for phasing our observations correctly, and for
relating them to a known ephemeris, but the difference between
BJD$_{\rm TDB}$ and HJD$_{\rm TT}$ is negligible compared to the {\it
  Kepler} exposure time.

\subsection{MAXI}

We also used X-ray monitoring data from the Gas Slit Camera (GSC) on
of the Monitor of All Sky Image \citep[MAXI;]{Matsuoka:2009a} on the
{\it International Space Station (ISS)}.  The GSC has a
$160^{\circ}\times1.5^{\circ}$ slit camera mounted at a fixed
orientation pointing away from the Earth.  Targets are usually
detected once per 92~min {\it ISS} orbit.  Transit times across the
slit are 45--120~s, so source fluxes are originally obtained with a
comparable timing precision to {\it Kepler}.  Since public light
curves are reported in observation period times we instead obtained
scan times from the {\it MAXI} team (T.  Mihara, 2015, private
communication) corresponding to the time during slit transit of
maximum transmission (i.e. effective mid-exposure).  These times are
reported in coordinated universal time modified Julian dates
(MJD$_{\rm UTC}$), where ${\rm MJD}={\rm JD} - 2\,400\,000.5$.  We
have converted MJD$_{\rm UTC}$ to HJD$_{\rm TT}$, consistent with our
{\it Fermi} light curves and within four seconds of the
BJD$_{\rm TDB}$ used with the {\it Kepler} light curve.  This should
result in tighter correlations between X-ray and optical flux and
sharper cross-correlation functions than earlier works using public
data products.  The GSC has a nominal energy range of 2--30~keV.  The
standard bandpasses which we used are defined as 2--4~keV, 4--10~keV,
and 10--20~keV, as well as a combined 2--20~keV band.

\section{Orbital Modulation}
\label{OrbitalSection}

We expect to find an orbital modulation of amplitude $\sim0.1$~mag in
the {\it Kepler} data as seen in previous studies described in
Section~\ref{IntroSection}. Fitting a sine wave to the data we measure
a photometric period of $P=0.787\,47(72)$~d, where the uncertainty is
estimated by bootstrap resampling as described in
Section~\ref{KeplerSection}.  This is comparable to other estimates
obtained from archival photometry
\citep{Gottlieb:1975a,Hynes:2012a,Galloway:2014a}, and consistent
within errors with the most precise spectroscopic period of
$P_{\rm orb} = 0.787\,311\,4(5)$~d measured using Bowen emission lines
\citep{Galloway:2014a}.  For the remainder of this work we will adopt
the latter period.

\begin{figure}
\epsfig{width=3.5in,file=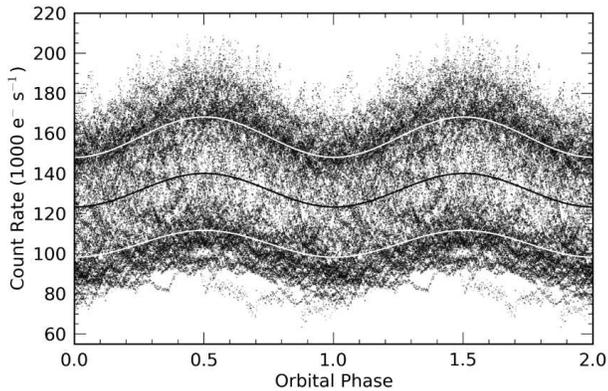}
\caption{{\it Kepler} data folded on the orbital period of
  \citet{Galloway:2014a}.  The solid black line is a sinusoidal fit to
  the raw data.  Solid white lines are the same fit scaled to match
  the two dominant groups of points. Note the distinct bimodal
  behaviour in optical brightness, which we are calling high and low
  states and also that both states apparently share the same
  modulation.}
\label{FoldFig}
\end{figure}

In Fig.~\ref{FoldFig} we show the {\it Kepler} data folded on
Galloway's period.  We also show a sinusoidal fit to the full data set.
The sine wave appears to describe the modulation fairly accurately; we
will discuss possible deviations from a pure sine wave shortly.  The
phase-offset with respect to the $T_0$ defined by
\citet{Galloway:2014a} is only 0.007 and the full-amplitude is
0.14~mag, comparable to earlier observations.  We derive an updated
time of photometric minimum as $T_0=2\,456\,932.7386 \pm 0.0016$
(BJD$_{\rm TDB}$) with the error estimated by bootstrap resampling.
For comparison, the ephemeris of \citet{Galloway:2014a} predicts a
time of minimum of $T_0=2\,456\,932.7430 \pm 0.0019$
(HJD$_{\rm UTC}$).  Accounting for the difference between time systems
this is a difference of 0.0052~d, or 0.007 in phase, as noted above.
Combining the errors on the two measures, this is a 2.1-$\sigma$
difference.  We conclude that the spectroscopic and photometric phases
are coincident to a high precision, better than 1~percent of an
orbit. The difference does not have high statistical significance.

It can be clearly seen in Figure~\ref{FoldFig} that the data follow a
bimodal distribution, as noted by \citet{Scaringi:2015a} and
\citet{Hakala:2015a}, and comparable to what was found by earlier
studies.  As shown in Fig.~\ref{FoldFig}, the fitted mean modulation
appears to separate the high and low state data quite well.  We could
use this dividing line to separate the data into high and low state
samples more cleanly than was possible ignoring the modulation. In
practice, the mean flux, as used by \citet{Hakala:2015a}, or the mean
modulation, as shown in Figure~\ref{FoldFig}, have the undesirable
property that the adopted division between states depends on the
relative time spent in each state, so we adopt a somewhat different
procedure. We instead detrend the light curve by dividing by a sine
wave with the same fractional amplitude but unit mean
($1 - A \cos(2\pi\phi)$, where $A$ is the fractional half-amplitude
and $\phi$ is the orbital phase), and then define the state boundary
at 140\,000\,e$^{-1}$~s$^{-1}$ after detrending. The choice of the
latter is motivated by examination of the optical versus X-ray
relations in Section~\ref{FluxDiagramSection} and we will defer more
detailed discussion until that point. With this prescription, the
state classification of a given point is independent of the relative
time spent in each state. Throughout this work we will refer to the
two optical states as the high state and low state, and this
terminology should not be confused with the high and low states of
black hole LMXBs (also known as thermal dominant and hard states
respectively). These are not used in discussing Z sources, so there
should be no confusion in this work.

\begin{figure}
\epsfig{width=3.5in,file=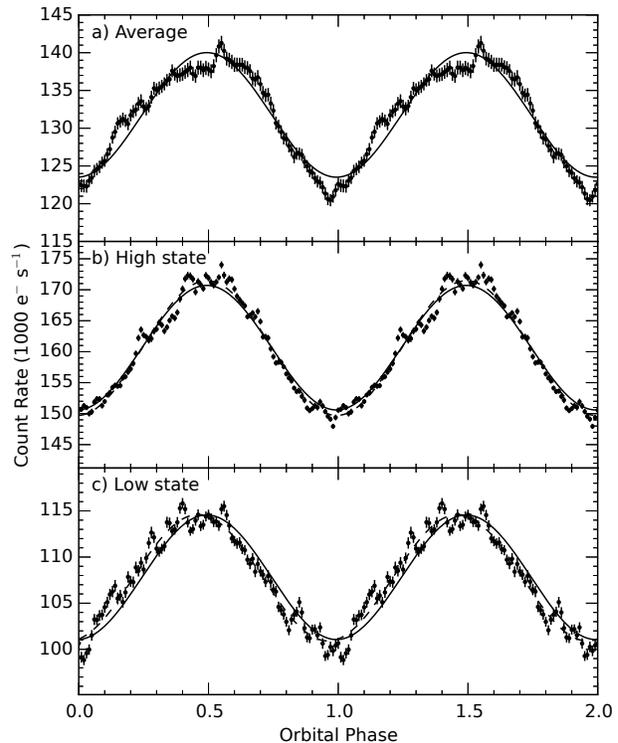}
\caption{Phase-binned data compared to sinusoidal fits.  Solid lines
  are the fit to the full data set, rescaled to match the mean flux in
  each panel.  The dashed lines in panels b) and c) are fits to the
  data in those panels only. Note that the dashed lines are very close
  to the solid line and indistinguishable in some places and that the
  apparent deviations from a sine wave are not statistically significant.}
\label{FitFig}
\end{figure}

We show in Figure~\ref{FitFig} a comparison of phase-binned
light curves, both the average of all the data and separate averages
of high and low state data, along with the sinusoidal fit to the whole
unbinned data set and fits to the individual binned samples.  A
sinusoidal fit is to be expected for a low inclination view of a
heated donor; the light curve becomes progressively non-sinusoidal at
higher inclinations, but at low inclinations ($\la 50^{\circ}$) is
almost indistinguishable from a sine wave.  We see that all the curves
are quite close to sinusoidal, as expected given the inclination
estimates (see Section~\ref{IntroSection}), and that deviations are
comparable to the amplitude of residual variability that has not
averaged out.

There is no clear difference in amplitude between high and low states.
Formally we measure a full amplitude of $0.147\pm0.012$~mag in the low
state and $0.151\pm0.008$~mag in the high state.  The low state does
suggest a small shift to earlier phases.  We measure the phase offset
between high and low state to be $0.032\pm0.024$, with the low state
data leading the high state.  If real, this might suggest either
changes in the light distribution on the secondary star, or changes in
disc asymmetry between the states. It is not significant at the
2-$\sigma$ level, however, and so may be an artefact of the random
realisation of aperiodic variability.

Examining the folded light curves, there appear to be apparent
deviations from a sinusoidal light curve in the average light curve,
and in particular it is noticeably more V-shaped at minimum than a
pure sine wave. This is reminiscent of a grazing partial eclipse of
the accretion disc, contrary to all the evidence for a low inclination
(Section~\ref{IntroSection}).  We note, however, that the amplitude of
fluctuations is comparable at other phases (e.g.\ near maximum). We
also tried folding subsets of the data taking either the first and
second 40~d period, or alternating odd and even two-day segments.  In
neither case was the effect seen in both subsets of the data, and
these subsets show quite large deviations due to variability that has
not averaged out.  The eclipse-like feature, and indeed all the
deviations from a sinusoidal form, appear therefore to be coincidental
and not significant.  In summary, we find no significant deviations
from a sinusoidal modulation, and no compelling differences in phase
or amplitude of the modulation between the low and high states. We
note that this conclusion is a little different to that drawn by
\citet{Hakala:2015a}. While they conclude that the amplitude is lower
in the low state, they express it relative to the mean overall flux,
rather than the mean low state flux. The folded light curves they show
do show approximately constant {\em fractional} amplitude when
compared to the mean flux of each state and this is consistent with
our finding of equal amplitudes in magnitude. \citet{Hakala:2015a}
also suggest that there are changes in morphology between states. As
argued above, we believe that these are non-repeatable artefacts
caused by not sufficiently averaging out variability. Since our state
division is defined differently to theirs, our folded light curves
will not average over variability in the same way, and so would be
expected to look somewhat different. In fact, by using a fixed rather
than phase-dependent division between states, \citet{Hakala:2015a}
will mistakenly assign some data at maximum light (phase 0.5) to the
high state instead of the low state, and vice versa at minimum light
(phase 0.0). This may have the effect of distorting their
state-dependent light curves leading to an underestimate of the
derived modulation amplitudes.

\begin{figure}
\epsfig{width=3.5in,file=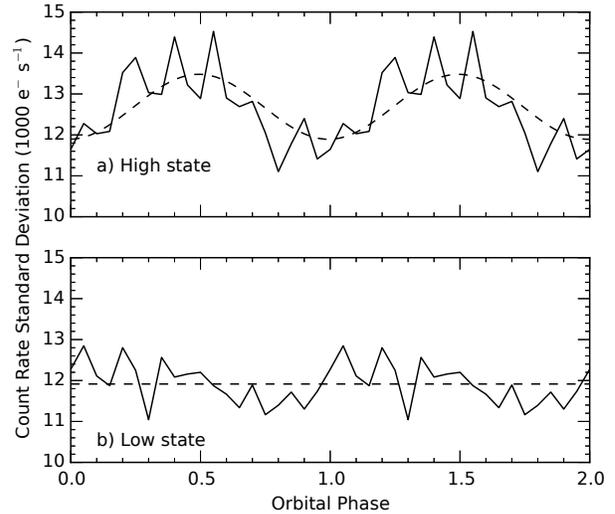}
\caption{Standard deviations within phase-binned {\it Kepler} data as
  a function of orbital phase and state. During the low state, the
  variability is essentially constant throughout the orbit, whereas
  during the high state, Sco~X-1 is more variable around phase 0.5
  (when the companion star is viewed behind the neutron star).}
\label{FlickerFig}
\end{figure}

We can also examine the standard deviation of the folded light curves
for the high and low states.  We show this in Figure~\ref{FlickerFig},
with a coarser binning than used for the mean light curve due to the
lower signal-to-noise ratio in the standard deviations than in the
mean light.  Surprisingly, the variability does appear to have a
different phase dependence in the two states, even though the mean
light does not.  We find that in the high state the variability is
modulated in the same way as the mean light, whereas in the low state
we see little if any modulation.  Put another way, in the low state
non-orbital variability has a constant absolute standard deviation (in
flux), whereas in the high state it has a constant fractional standard
deviation.  To investigate the significance of this result we repeated
the analysis using bootstrap resampling as done earlier in this
section. We quantify the modulation by fitting a sine wave of fixed
period and phased the same as the mean modulation, leaving the
amplitude and zero point as free parameters. We find from 1000 samples
that the modulation amplitude in the high state is
$7.0\pm3.6$~percent, while in the low state it is
$-0.4\pm3.4$~percent. This leaves the difference in behavior
suggestive but inconclusive.

If the difference in modulation of the variability between the normal
and flaring branch is real, it suggests that in the high state, the
variability is imprinted on all of the emission by a common source.
Since the high state appears to be associated with the X-ray FB where
short term X-ray-optical correlations are seen
\citep[][Sections~\ref{LightcurveSection},
\ref{FluxDiagramSection}]{MunozDarias:2007a,Britt:2013a,Scaringi:2015a},
the likely mechanism is reprocessing by both the disc and secondary
star of X-ray variations.  In the low state, the mean light curve
indicates that irradiation of the companion is still occurring, but
the aperiodic variability appears to be dominated by the disc, since
it does not modulate strongly on the orbital period.  This is
consistent with the lack of clear, positive X-ray-optical correlations
in the NB which appears to correspond to the low state
\citep[][Sections~\ref{LightcurveSection},
\ref{FluxDiagramSection}]{Britt:2013a,Scaringi:2015a}.

\section{Detrended light curves}
\label{LightcurveSection}

\begin{figure*}
\includegraphics[width=7.7in, trim=0.7in 0.0in 0.0in 0.0in, clip]{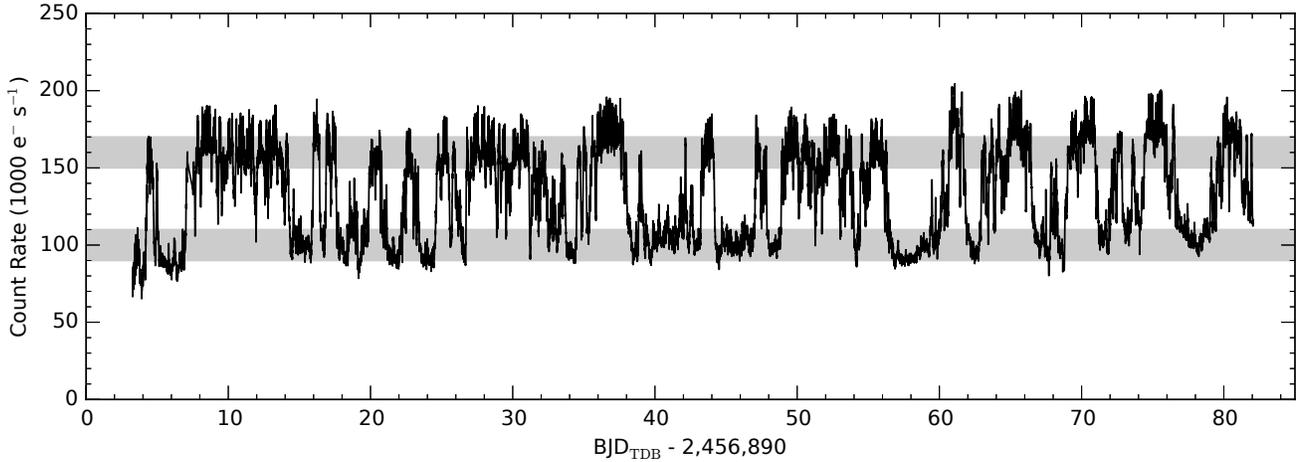}
\caption{The whole {\it Kepler} light curve after removing the orbital
  modulations.  Upper and lower bands correspond to fluxes around the
  two peaks of the histogram in Figure~\ref{HistFig}. This is the
  first optical light curve to show the structure of the transitions
  between high and low states, with characteristic times in each state lasting for
  typically two to five days.}
\label{LongLCFig}
\end{figure*}

In Section~\ref{OrbitalSection} we fitted the sinusoidal orbital
modulation and produced a prescription for detrending the light curve.
This can be expected to produce simpler light curves, and also cleaner
division into states in flux histograms.  We show the whole detrended
light curve in Figure~\ref{LongLCFig}.  The bimodal distribution is
clear even here, with irregularly alternating high and low states and
superposed faster variability.  The low state shows mostly brightening
episodes, while the high state shows both fast flares and slower
dips. We will reserve the term `flares' to refer to the discrete, fast
variations in the high state (which corresponds to the X-ray FB as
noted above), and refer to the slower variations as brightenings or
dips. The latter resemble incomplete or failed state transitions.

\begin{figure}
\epsfig{width=3.5in,file=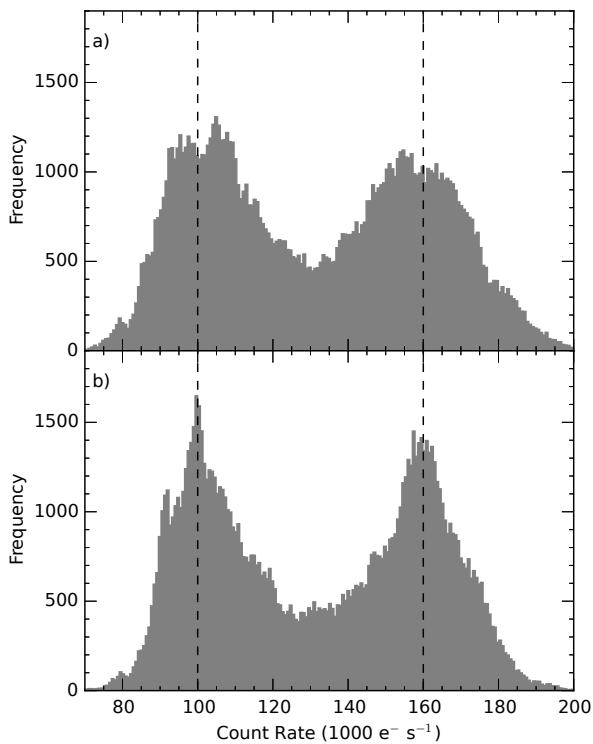}
\caption{Flux histograms before (a) and after (b) removing the orbital
  modulation.  Dotted lines are placed at 100\,000~e$^{-}$~s$^{-1}$
  and 160\,000~e$^{-}$~s$^{-1}$ to mark the two main peaks. The
  histogram clearly shows that Sco~X-1 is usually in either a fairly
  narrow high state or low state, with the fill in the middle caused
  by transitions (or failed transitions) between the two states.}
\label{HistFig}
\end{figure}

We show histograms before and after detrending the orbital modulation
in Figure~\ref{HistFig}.  The improvement is pronounced near the
peaks, but some overlap between the states remains.  This is to be
expected, as state transitions are well resolved, and intrinsic
variability within the states may also result in some overlap.

\begin{figure}
\includegraphics[width=3.4in, trim=0.0in 1in 0.0in 1in, clip]{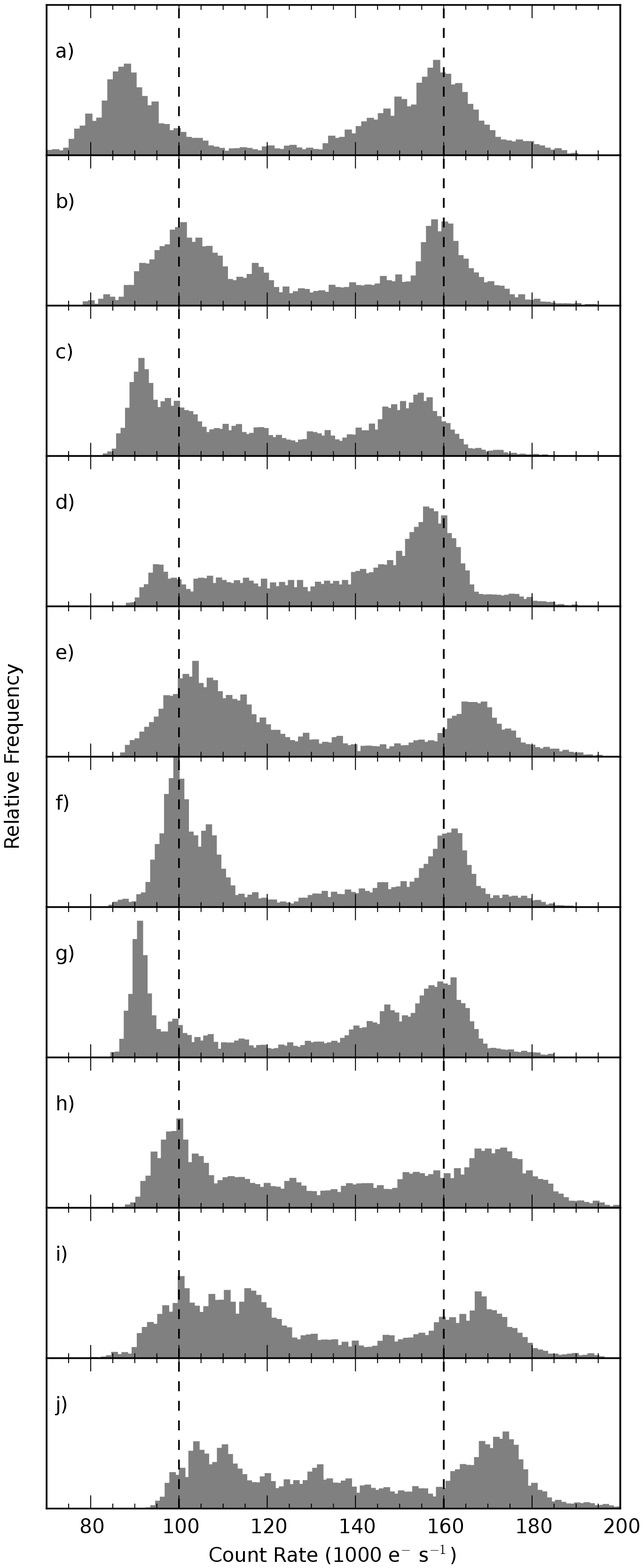}
\caption{Histograms of ten equal subsets of the light curve after
  removing the orbital modulation.  Dotted lines are as indicated on
  Figure~\ref{HistFig}. For all the Kepler data, Sco~X-1 only shows two states, with
 these moving around somewhat in brightness.}
\label{MultiHistFig}
\end{figure}

A single long observation such as provided by {\it Kepler} is ideal
for putting historical studies of flux histograms into context.  These
studies showed a variety of bimodal and trimodal histograms, with
variations from year to year.  These can be interpreted in two ways;
either there are three distinct optical states, of which sometimes
only two are seen, or there are only two states which change in
brightness enough that incompletely sampled data might show them
appearing as three states.  In Figure~\ref{MultiHistFig} we show the
histograms of the {\it Kepler} data subdivided into ten consecutive
segments each of about eight days duration, long enough to sample a
range of states.  We do see variations in the flux level of the high
and low states, and the separation of the two states also clearly
varies.  At no point do we see a clear trimodal distribution. However,
the variations we see over the course of about three months are not
large.  The lowest peak of the low state corresponds to about 0.15~mag
below its average value and the highest peak of the high state is
about 0.1~mag above the mean.  Historical histograms sometimes show
trimodal distributions with peak separations $\sim0.5$~mag within a
single month.  This could not be reproduced by any reasonable sampling
of the {\it Kepler} data.  This analysis then requires that the
historical behaviour be different to that seen during the {\it Kepler}
observation.  Either the flux levels of the states were more variable,
or a third state not seen by {\it Kepler} was present.  Since only the
NB and FB can be clearly identified in the MAXI Z diagram
(Section~\ref{FluxDiagramSection}) and these can be associated with
{\it Kepler} low and high states, it is possible that the missing
third state is associated with the HB \citep[c.f.][]{Vrtilek:1991a}.
The subdivided histograms also show that the infilling between the
peaks in Figure~\ref{HistFig} may be largely a consequence of the
movement of the peaks.  Figure~\ref{MultiHistFig}a, for example shows
negligible infilling between the peaks and may provide the best
indication of the intrinsic flux distributions of the two states.
Both peaks show asymmetry of character similar to that suggested by
\citet{Scaringi:2015a}, with the high state showing a low tail and the
low state showing a high tail. These asymmetries presumably arise from
the intrinsic asymmetry of the failed transitions (brightenings in the
low state and dips in the high state.)

\citet{Scaringi:2015a} used MAXI colours to identify FB and NB epochs.
They found that the high state was usually associated with the FB and
that the low state corresponded to the NB.  They also found that the
optical flux was a better predictor of X-ray branch than X-ray fluxes
alone.  Since we cannot classify states based on colour with GBM data,
for this work we will use the optical fluxes to divide data into high
states and low state using a dividing line at
140\,000~e$^{-}$~s$^{-1}$ (Section~\ref{FluxDiagramSection}). These
provide a proxy for the X-ray FB and NB, but we should not assume that
they are completely synonymous. We will examine the association
between optical and X-ray states more carefully in
Section~\ref{FluxDiagramSection}.

\section{Discrete Statistical Analyses}

\subsection{High State Durations}
\label{HighStateDurationSection}

The {\it Kepler} light-curve represents an unprecedented source of
information on the statistical distribution of source states which is
relatively free of completeness effects since it contains very few
gaps. We have so far considered distributions of continuously variable
fluxes. We now move to consider discrete states, transitions, and
events. We begin by considering typical durations of the high
states. We note that definition of high states is relatively
straightforward, since they are characterized by a quite flat plateau
with signature flares. An analogous distribution of low states would
not be possible, as the low state merges seamlessly into transitions.

\begin{figure}
\epsfig{width=3.5in,file=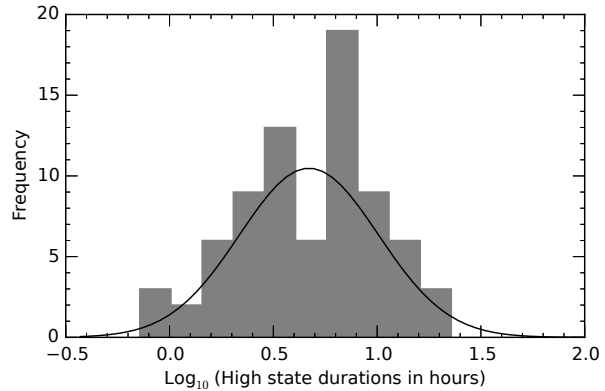}
\caption{Distribution of durations of uninterrupted high
  state episodes. The fit is a log-normal distribution described in
  the text.}
\label{HighStateDurationFig}
\end{figure}

We define a continuous high state as a period when the source showed a
relatively flat plateau which included flares. A continuous high state
is terminated by any clear dips below the plateau level. In some
cases, e.g. around days 7--14 in Figure~\ref{LongLCFig}, this breaks a
relatively long high state into multiple shorter episodes. We believe
this is an appropriate definition, as the high state is interrupted by
an attempt, albeit a failed one, to transition to the low state. With
this definition we identify 77 high state periods of durations
0.7--22.8~hours.  There is a clear preference for quite short
durations, with longer uninterrupted periods becoming increasingly
improbable. The distribution can be fit by a log-normal distribution
with shape parameter 0.77 and scale parameter 4.7~hours. That is to
say, the logarithms of high state durations follow a Gaussian
distribution with standard deviation (of the logarithms) of 0.77 and
median duration 4.7~hours. The fit is acceptable to a
Kolmogorov-Smirnov test with a p-value of 0.45. We show the resulting
distribution of durations together with the fitted distribution in
Figure~\ref{HighStateDurationFig}. A log-normal distribution arises
when results are a product of multiplication of a series of
independent processes. Such a model for X-ray binary variability is
indeed now favored in the form of the propagating perturbation model
in which local mass transfer fluctuations at progressively smaller
radii imprint their signature multiplicatively on the overall mass
transfer rate and hence source brightness
\citep{Lyubarskii:1997a}. This does appear to lead to a log-normal
distribution of fluxes \citep{Uttley:2005a,Gandhi:2009a}. If our high
states reflect the length of time where the flux exceeds a threshold
level, then their duration distribution would be related to the flux
distribution, and so a log-normal duration distribution might
naturally be expected.

We can also take a broader definition of high states, as we see
extended periods where the high state is preferred, even if brief
partial or complete transitions back to the low state occur. The
longest such episodes last for about a week, and are followed by an
extended low or low/transition period. We will discuss the possible
significance of this in Section~\ref{DiscussionSection}.

\subsection{Transition Time-scales}

It is quite striking that there is a similarity of time-scales among
all transitions between the high and low states. This applies to full
transitions, brightenings in the low states, and dips in the high
states. In fact, if the source brightness is plotted on a magnitude
scale, these transitions appear to be near-linear with a modest (but
clear) range of gradients. This suggests an exponential
behavior that can be characterized by an e-folding time, defining a
time-scale for the process mediating the optical transition.

We compile measurements of all such transitions where a near-monotonic
rise or decline segment of at least 10 percent in flux could be
identified. We find no convincing dependence of the transition
time-scale on either the mean brightness or the amplitude of the
transition, suggesting that large state transitions, low state
brightenings, and high state dips are all intimately connected. There
does appear to be some difference between rise and decay time-scales,
with slower decays more prevalent than slow rises, but the
distributions of rise and decay time-scales are strongly
overlapping. A Kolmogorov-Smirnov test rejects a common distribution
for the two at a significance of 99.9~percent. Both rise and decay
time-scale distributions are consistent with log-normal distributions
with shape parameters of 0.37 and 0.43 and scale parameters (median
time-scales) of 2.9~hours and 3.4~hours respectively. The fits are
both acceptable to Kolmogorov-Smirnov tests with p-values of 0.96 and
0.17 respectively. The time-scales quoted are e-folding time-scales,
not actual transition times. The ratio of optical fluxes between
typical high and low states is 1.6 (Figure~\ref{HistFig}), so the
characteristic times to execute a full rising and falling transition
are 1.7 and 2.0~hours respectively.

\begin{figure}
\epsfig{width=3.5in,file=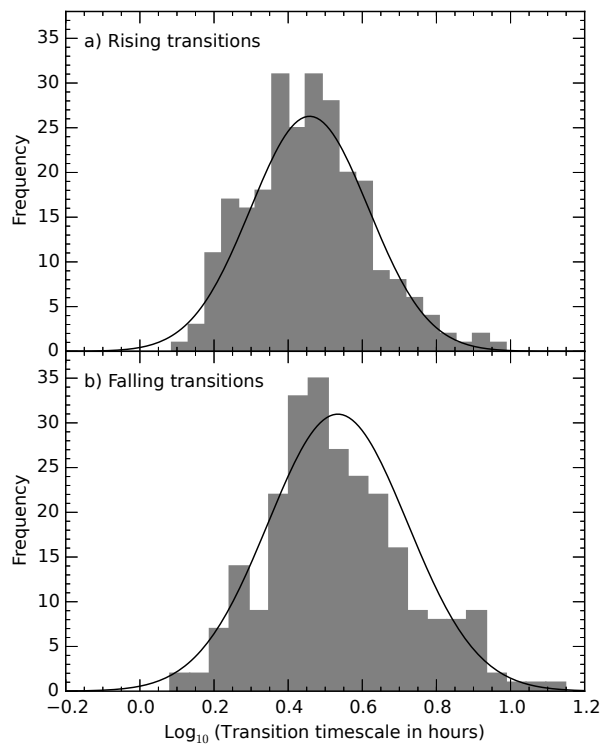}
\caption{E-folding time-scales of transitions between high and low
  states, including failed transitions. The smooth lines are
  log-normal fits to the distributions described in the text.}
\label{TransitionTimescaleFig}
\end{figure}

\subsection{High State Flare Characteristics}
\label{FlareSection}

\begin{figure}
\epsfig{width=3.5in,file=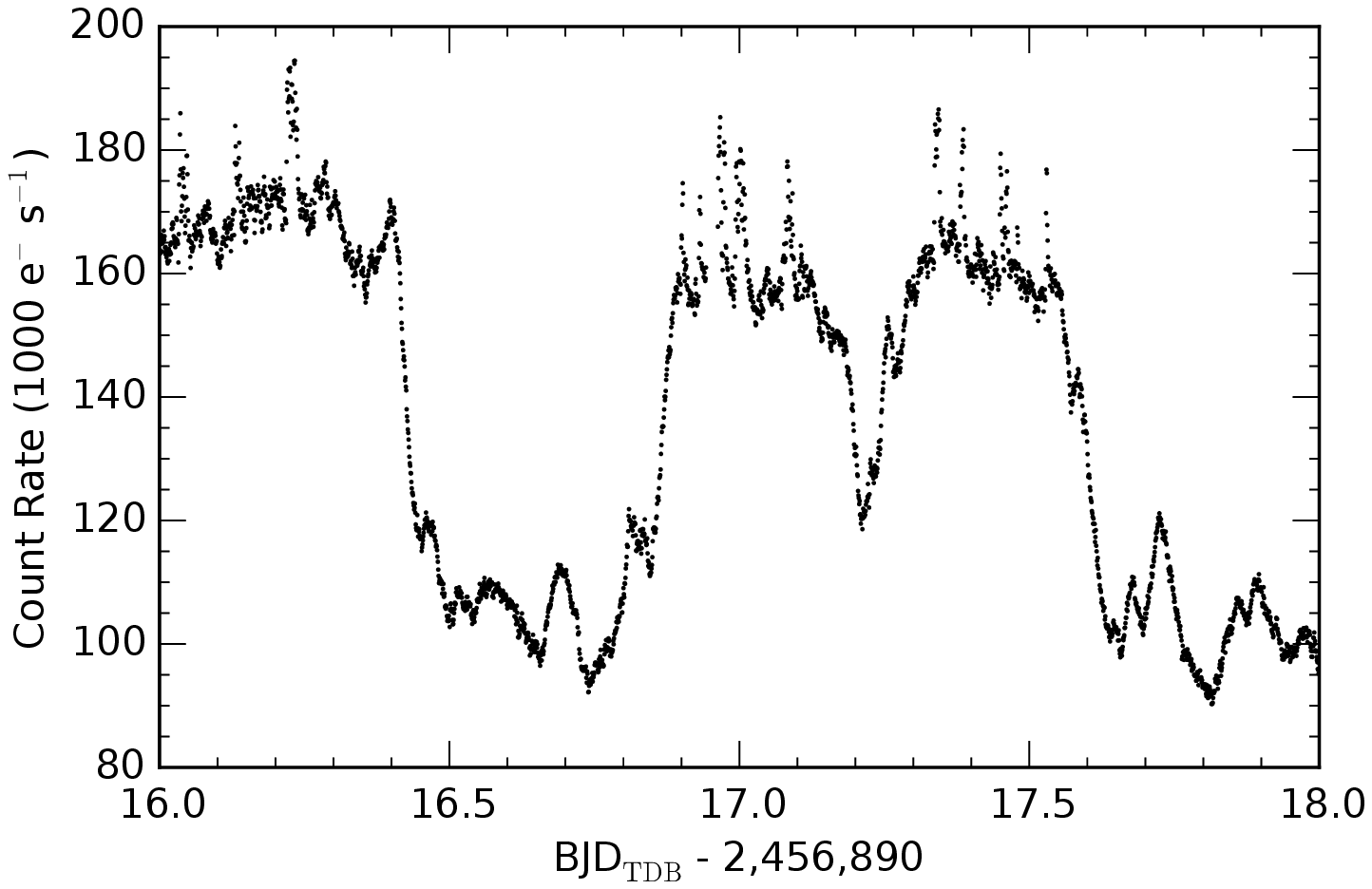}
\epsfig{width=3.5in,file=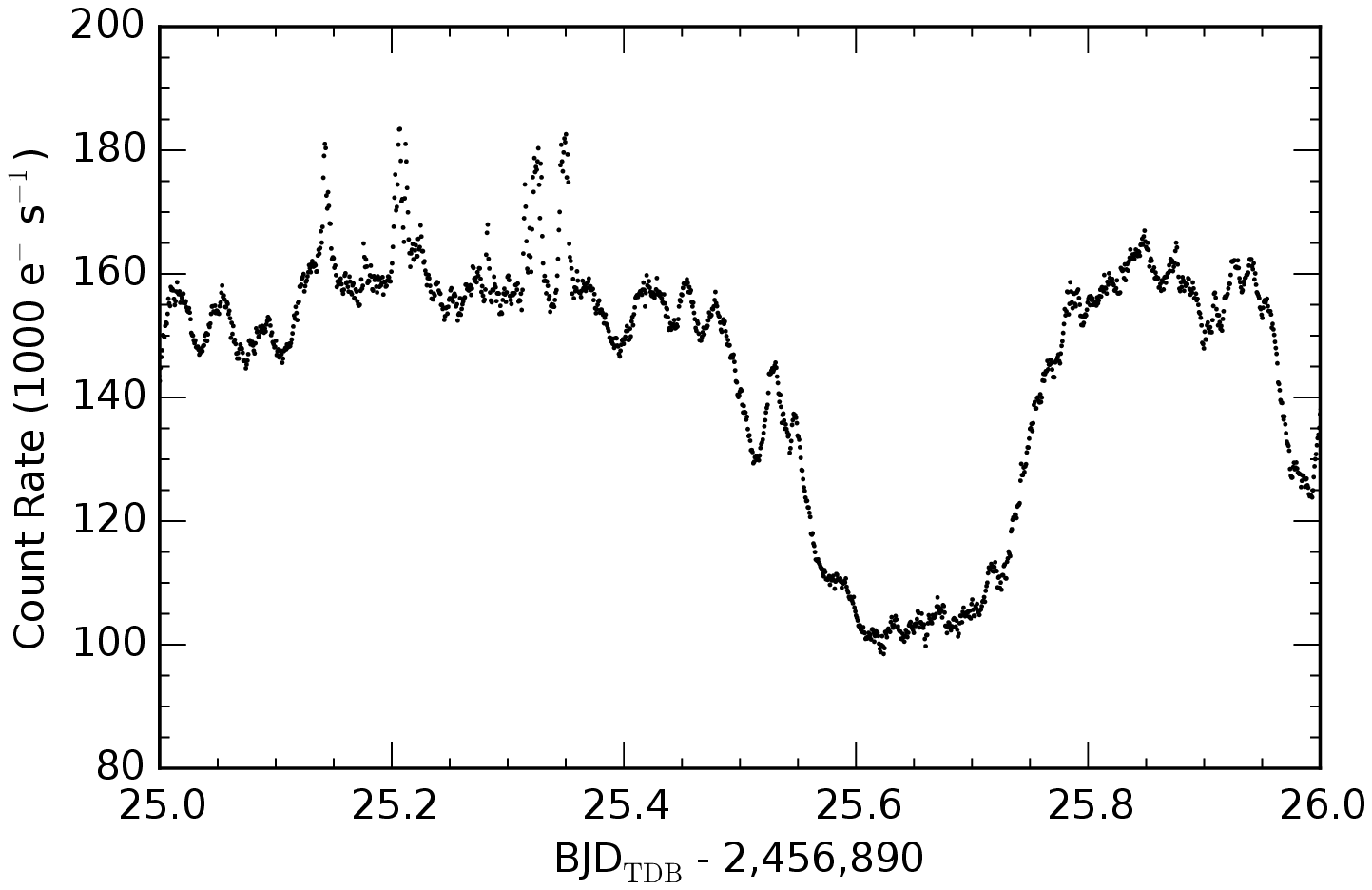}
\epsfig{width=3.5in,file=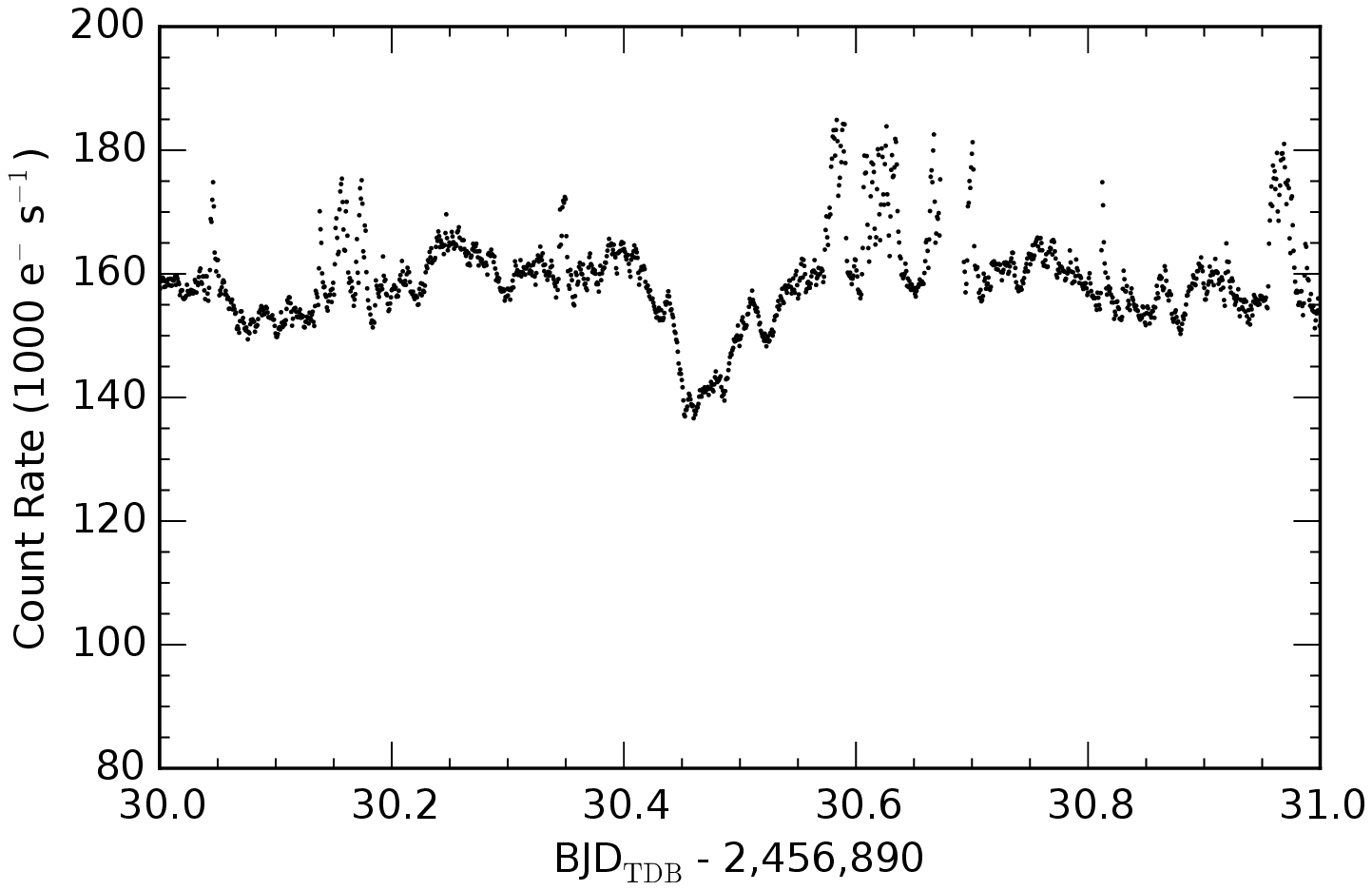}
\caption{Examples of high state {\it Kepler} light curves. When
  Sco~X-1 is in its high state, it frequently displays fast flares,
  and these fast flares are only seen during the high state.}
\label{XXnew}
\end{figure}

While Sco~X-1 is on the FB it shows strong X-ray flares. We show some
examples of corresponding events seen in optical light in the high
state in Figure~\ref{XXnew}.  Detailed prior studies with simultaneous
X-ray and optical time series show that the optical light from flares
is a slightly delayed and smeared version of the X-ray flares,
pointing to the optical flare flux coming from X-rays reprocessed by
both the disc and the companion star (see review in
Section~\ref{IntroSection}).  Our {\it Kepler} program for Sco~X-1 now
has by far the longest and most completely sampled optical light
curve.  This long and steady coverage allows for the demographics of
the flares to be well-measured.  Detailed examination of the full
light curve shows that the optical flares are distinct events with a
fast rise, a roughly symmetric light curve shape, durations of
5--20~min, all with nearly the same amplitude, and {\it only}
occurring when Sco~X-1 is in its nearly-flat high state.

The optical flares are distinct short-duration brightenings from the
optical high state that always have a sudden start, a sharp peak, and
a sudden stop.  We see no case where there is any prolonged pre-flare
or post-flare emission.  To get some statistics on flare durations and
rise/fall times, we have gone through the light curve and picked out
by eye the bins that are the start, peak, and end of each optical
flare.  From these points, we can calculate the rise time, the fall
time, the duration, and the amplitude of the peak of each flare.  The
histogram of the flare durations shows a peak at 9.4~min, with the
central 67~percent of the flares with durations between 8.3 and
16~min, and with the central 90~percent of the flares between 5 and
31~min. If the longer duration flares are actually two or more flares
superposed at nearly the same time which we could not resolve into
separate flares, as appears to be the case, then the real upper limit
for the duration of a single flare is roughly 20~min.  Flares with
significant flux and durations from 2--5~min would have been easily
identified, so the deficit of $<$5~min duration flares appears to be
real.  Indeed, all the flares identified as having durations $<$5~min
are tightly bounded by other flares such that the times tagged for the
start and end are high above the usual floor. These short durations
are therefore more like full-width-half-maximum durations, so the
$<$5~min durations appear to be an artefact of flare bunching.  There
is no preference for rise times to be either faster or slower than
fall times.  So the optical flare light curves are distinct with steep
rises, sharply isolated from the earlier and later variations,
5--20~min durations, and roughly symmetric shapes. We note that
\citet{Scaringi:2015a} analyzed the {\it Kepler} power spectrum in the
high state and identified two structures, a red noise power law and a
bump around $10^{-3}$~Hz, corresponding to time-scales of
15--20~min. This bump is presumably associated with the rapid flares
which are distinct from the underlying flickering of the plateau
level.

\begin{figure}
\epsfig{width=3.5in,file=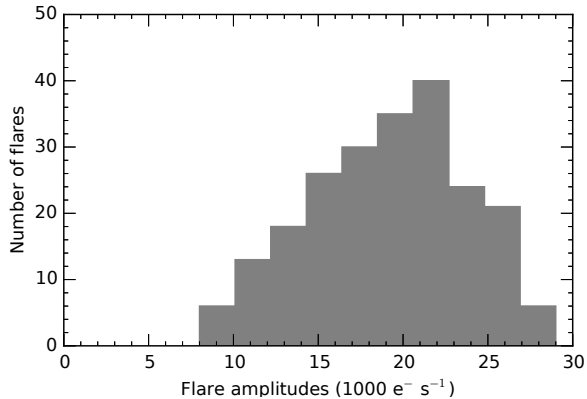}
\caption{Distribution of high state flare amplitudes. The amplitude of
  the fast high-state flares varies only over a fairly restricted
  range with a quite sharp upper cutoff.}
\label{YYnew}
\end{figure}

In looking at the light curve with Sco~X-1 in its optical high state,
it apparent that all the optical flares have similar peak amplitudes
above the non-flare level.  This is apparent in the three panels of
Figure~\ref{XXnew} and also in Figure~\ref{SampleHiFig}.  To quantify
this, we have constructed a histogram of flare amplitudes (see
previous paragraph) from the peak level to the base level at the start
or end of the flare.  Our peak amplitude histogram (see
Figure~\ref{YYnew}) shows a peak around 20\,000~e$^{-}$~s$^{-1}$, with
a minimum duration of 8\,000~e$^{-}$~s$^{-1}$ and a sharp upper limit
of about 27\,000~e$^{-}$~s$^{-1}$. Our light curve time resolution is
59~s, which is longer than the usual time at peak, so our peak fluxes
will usually be underestimates by variable fractions, and this will
lead to a moderate intrinsic scatter in the peak amplitudes.  Another
source of moderate intrinsic scatter for our measured amplitudes comes
from partial overlaps for a small fraction of the flares, where the
rise or decline of one flare adds to the peak flux of another flare,
creating an apparently larger amplitude.  The histogram shows sharp
upper and lower limits on the flare amplitude.  The upper limit is
real and we can think of no artefacts or biases that would eliminate
flares with amplitude $>$27\,000~e$^{-}$~s$^{-1}$.  The real sharpness
of the lower limit is problematic, because we cannot readily quantify
the efficiency of our selection for low amplitude flares.  Looking at
the light curves, we see few events that could be claimed to be flares
with amplitude $\sim$8000~e$^{-}$~s$^{-1}$, but we do see variations
at the $\sim$2000~e$^{-}$~s$^{-1}$ level that might be low amplitude
fast optical flares.  Thus we think that the optical fast flares do
have a lower limit on their amplitude, but this conclusion is not of
high confidence.  From this analysis, it appears that the fast optical
flares are all roughly of the same amplitude
($\sim$20\,000~e$^{-}$~s$^{-1}$), and there is certainly a sharp upper
limit on flare amplitude ($\sim$27\,000~e$^{-}$~s$^{-1}$).  An obvious
interpretation is that the upper limit in the optical comes from a
maximum luminosity associated with the nuclear burning event on the
neutron star, much like in the case of X-ray bursters. This is
consistent with the model of \citet{Church:2012a} for the FB where
unstable nuclear burning is providing part of the luminosity of
flares.

\begin{figure}
\epsfig{width=3.5in,file=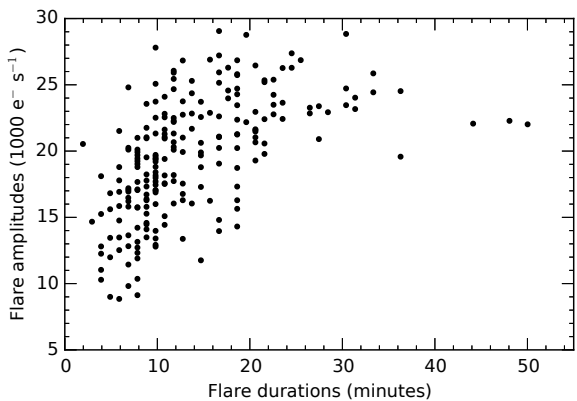}
\caption{Relationship between durations and amplitudes of high state
  flares. A significant and positive correlation shows that the larger amplitude
 flares have a strong tendency to be longer in duration. The flares
 with apparent duration longer than 30~min are likely just two or
 three flares spaced closely enough in time that we could not
 distinguish the individual flares.}
\label{AAnew}
\end{figure}

Both the flare amplitudes and the flare durations have relatively
small ranges.  Over those ranges, the amplitudes and durations are
positively and significantly correlated (see Figure~\ref{AAnew}).  The
existence of this correlation suggests that the flare amplitude is not
some constant `standard candle' (as for X-ray bursts), even though the
range of amplitudes is apparently fairly restricted.  We do not find
any significant correlation between the time between flares (either
before or after) and the flare amplitude or duration.  We do not have
a ready explanation as to why the durations and the amplitudes get
larger and smaller together over many flares.

The Sco~X-1 fast optical flares have a distinct morphology, and these
are definitely different from the slower brightenings and dips.  The
low state brightening events have durations from 0.02 to 0.10~d
(30--144~min) with no overlap in duration with the high-state fast
flares.  These brightening events all have a smooth light curves.
Their light curves are variously rounded or cuspy at the top, giving
the visual impression that the peaks are formed by failed transitions
between high and low states.  That is, if the system conditions that
determine the state are toggled over a short time interval, then dips
and brightenings will be made with the light curve variations being in
the transition region between high and low states, i.e. failed state
transitions.  We conclude that there are two morphologically distinct
types of brightening events in the Sco X-1 light curve, first the fast
optical flares associated with the fast X-ray flares, and second the
long-duration and smooth brightenings that are apparently failed
transitions between states.

The optical high state appears to be fairly constant plateau with a
flat light curve, albeit with the baseline flickering with the red
noise power spectrum noted above \citep{Scaringi:2015a}.  The fast
optical flares appear to be simple additions of the flare flux on top
of this plateau and are distinct from the flickering
behaviour. Another manifestation of this is the break in the high
state rms-flux relation shown by \citet{Scaringi:2015a}.  The
transitions to the low state make for quick fades from the
high-state-plateau as well as fast rises to the plateau.  Both the
flares and the transitions contribute substantially to the broadening
of the histograms of orbit-modulation-corrected flux for the
high-state.  We can define a pure-optical-high state as being when Sco
X-1 is optically bright, not rising or falling from a transition, and
not during a fast flare.  Sco~X-1 is in this state about a quarter of
the time.  During this pure-optical-high state, the flux is still
varying on fast and slow time-scales.  Thus, we see some small
amplitude variability that does not qualify as our fast optical
flares, and we see some variations up and down in the plateau level.
With all these superposed variations, nevertheless, the
pure-optical-high state appears to be essentially a nearly flat
plateau.

The fast optical flares {\it only} occur during the flat plateau of
the high state.  That is, we never see any fast flares in the low
state, in the transition state, during small dips in the high-state
plateau, or even on the edges when the transition state is merging
with the high state.  As already noted, and discussed further in
Section~\ref{FluxDiagramSection}, the high state is near synonymous
with the X-ray FB in the Z-diagram.  So we have equalities between
events and states from the X-ray and optical light curves, where the
X-ray fast flares are the same events as the optical fast flares, and
the X-ray FB is identical with the optical high state.  The optical
fast flares are caused by the X-ray flares produced on or near the
surface of the neutron star (perhaps from episodic nuclear burning),
and these are only seen during the optical high state, which is to say
that the X-ray flares only occur during the X-ray FB.  So the branch
on the Z-diagram is well named, and we are showing here with the long
{\it Kepler} data set that the other branches do not have flares.

\begin{figure}
\epsfig{width=3.5in,file=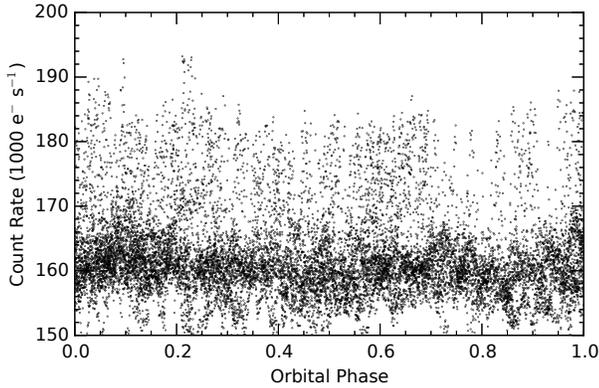}
\caption{The first 30~d of {\it Kepler} high state data folded on the
  orbital period.  The orbital modulation has been corrected,slow
  variations in the baseline level have been subtracted out and
  excursions towards the low state have been excluded. This figure
  shows the upper limit on the flare amplitude (with flare maxima
  between roughly 185\,000 and 192\,000~e$^{-}$~s$^{-1}$ above the
  baseline at 160\,000~e$^{-}$~s$^{-1}$).  This figure also shows that
  the flare amplitude (scaled to take out the orbital modulation) and
  frequency do not vary with orbital phase.}
\label{ZZnew}
\end{figure}

We have constructed a light curve of the optical high state with the
fast flares included and the transitions excluded, all with the
orbital modulation removed.  For this, we have further normalized out
the slow variations in the plateau level by picking out one or two
points a day as representative of the plateau variations and using
them to define a piecewise continuous light curve that can be
subtracted out.  The result is a nearly flattened light curve only of
the optical high state with the fast flares riding on top.  This light
curve has been folded on the known orbital period, as shown in
Figure~\ref{ZZnew} for the first 30~d.  This plot shows the sharp
upper cutoff of the flare amplitudes.  This also shows that this
plotted flare amplitude does not change substantially with orbital
phase.  Recall that this plot has already been corrected for the
orbital modulation by a multiplicative factor that varies sinusoidally
with phase so as to flatten the light curve.  Thus, the uncorrected
flare amplitude must scale with the mean optical light. This is a more
specific restatement of the finding in Section~\ref{OrbitalSection}
that in the high state, the standard deviation appears to be modulated
in the same way as the average light. The plot also shows that the
flare frequency does not vary with the orbital phase.

\section{Correlations with MAXI and Fermi GBM}
\label{FluxDiagramSection}

Early coordinated studies identified a pattern of correlations between
X-ray and optical flux \citep{Mook:1975a,Bradt:1975a,Canizares:1975a}.
At low optical brightnesses the X-ray emission appeared approximately
constant, but above a threshold optical brightness X-rays became
extremely variable and on average brighter.  \citet{McNamara:2003a}
refined the correlation using BATSE and ground-based optical fluxes,
and noted an inverse relation between optical and X-ray flux at low
optical brightnesses, but the data were sparse.
\citet{Scaringi:2015a} showed the correlations between {\it Kepler}
and MAXI exhibiting a huge improvement over earlier work, with the NB
anti-correlation clear and large X-ray variability in the FB.  For
comparison, \citet{OBrien:2004a} presented a comparable correlation
for Cyg~X-2, in which X-rays first increased then decreased with
optical flux as the Z diagram was traced from HB to FB. This is the
{\em opposite} of what is observed in Sco~X-1.

\begin{figure}
\epsfig{width=3.5in,file=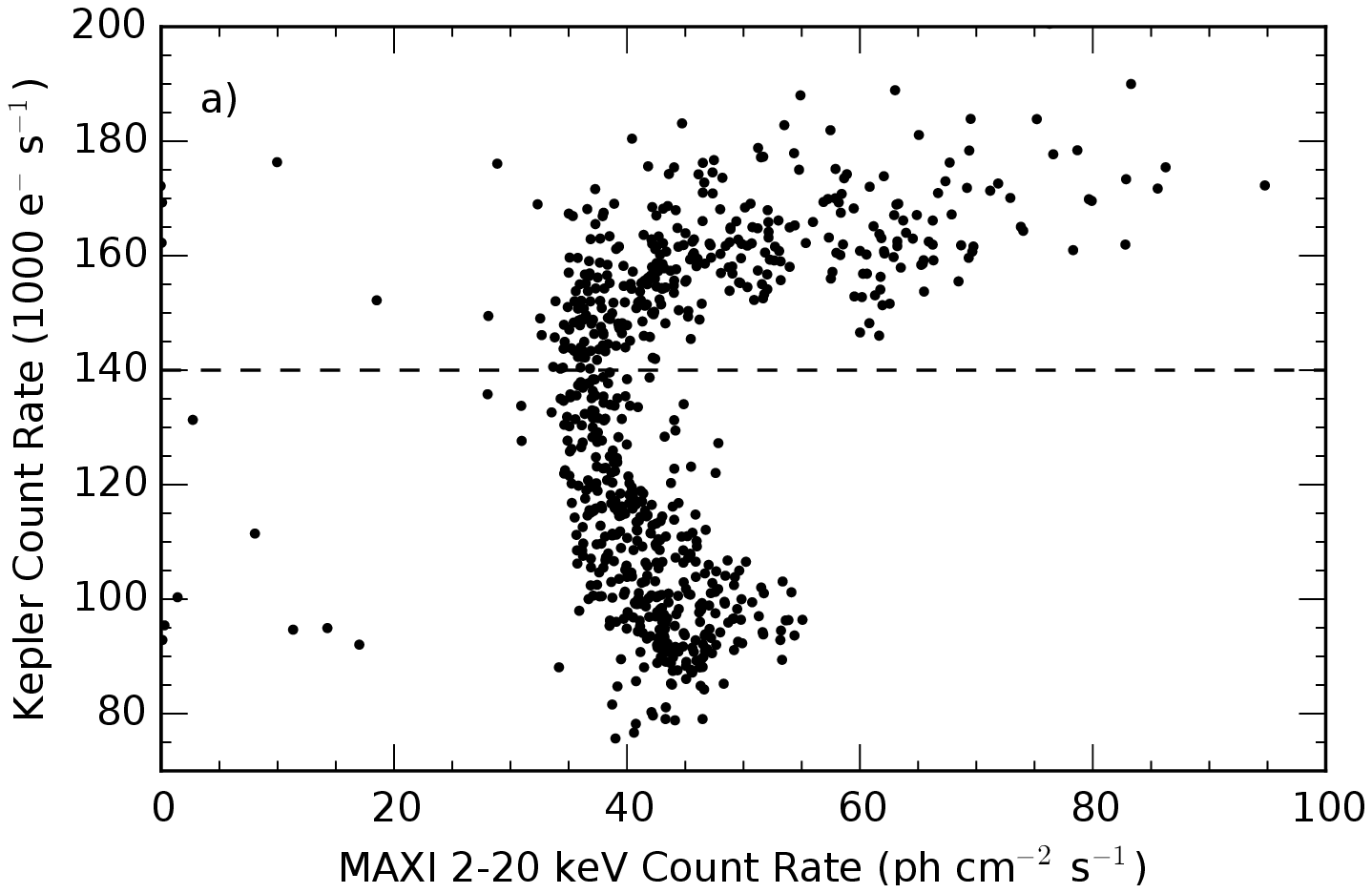}
\epsfig{width=3.5in,file=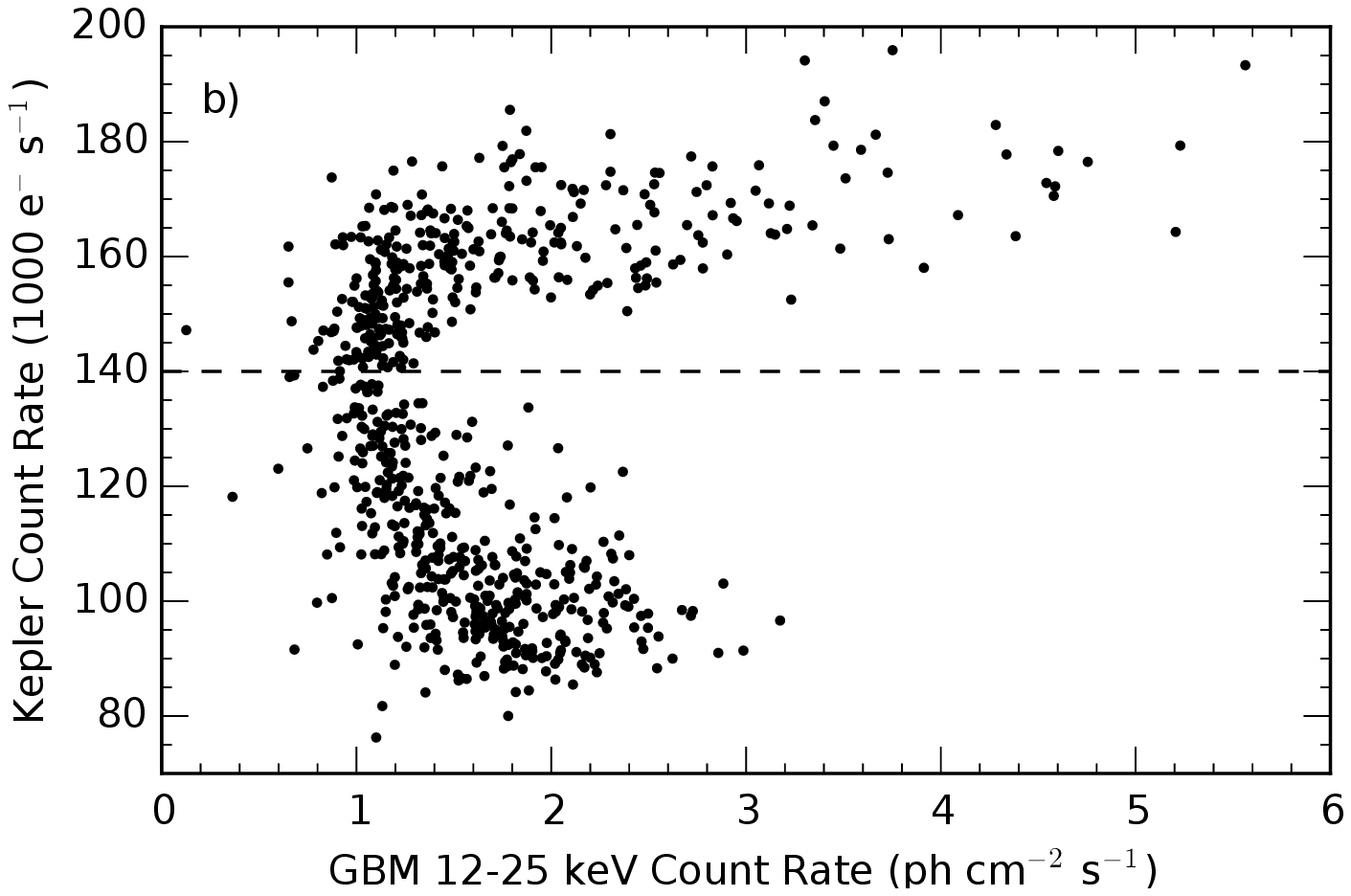}
\caption{{\it Kepler} vs. MAXI and GBM intensity-intensity diagrams.  The
  dashed line indicates our division between high and low
  states. This figure clearly shows the low state anti-correlation. It
  can be seen that the adopted dividing line corresponds to both the
  minimum hard X-ray flux, and the minimum X-ray variance.}
\label{IntensityFig}
\end{figure}

Our multi-wavelength data set includes two quite different subsets. The
{\it Kepler} optical data are near-continuously sampled with gaps
between exposures usually small compared to the exposure time. The GBM
and MAXI data are extremely sparsely sampled and consequently do not
represent well-resolved X-ray light curves. We can none the less
produce pairs of matched X-ray and optical points, either obtained
simultaneously, or for a given choice of lag, and these allow us to
examine correlations.  We show in Figure~\ref{IntensityFig} both the
MAXI vs. {\it Kepler} and GBM vs. {\it Kepler} relations.  By using
MAXI scan times and GBM occultation times we can match with the
optical more precisely than by using binned {\it Kepler} light curves.
In addition, by detrending the orbital modulation we can remove some
of the vertical scatter in {\it Kepler} data.
Figure~\ref{IntensityFig} then defines the best relations yet
available between optical and X-ray fluxes in Sco~X-1.

The diagrams can both be clearly delineated into two regions.  In the
high state, we see a small increase in baseline hard X-ray flux with
increasing optical brightness, and a large increase in the dispersion.
In the low state, there is an anti-correlation with the hard X-ray flux
increasing as the optical flux decreases.  The hard X-ray dispersion
also seems to increase again as optical flux decreases (by somewhat
more than expected for constant fractional variability).  In between
the two regimes we can define a transition region, for {\it Kepler}
fluxes around 140\,000~e$^{-1}$~s$^{-1}$, where both X-ray flux, and
X-ray variance seem to be minimized.  We can use this region to define
a cutoff to separate high and low states based on optical brightness
alone. It should be noted that this cutoff lies quite closely under
the high state in the diagram, and so it assigns much of the
transitionary data to the low-state. Since the general X-ray to
optical relation (i.e. an anti-correlation) is consistent below
140\,000~e$^{-1}$~s$^{-1}$ this may be the most meaningful division to
make, indicating that the low state behavior persists until the
transition to the high state is complete. This impression is also a
consequence of the much steeper dependence of the optical flux on
X-ray flux in the low-state than the high state.

\begin{figure}
\epsfig{width=3.5in,file=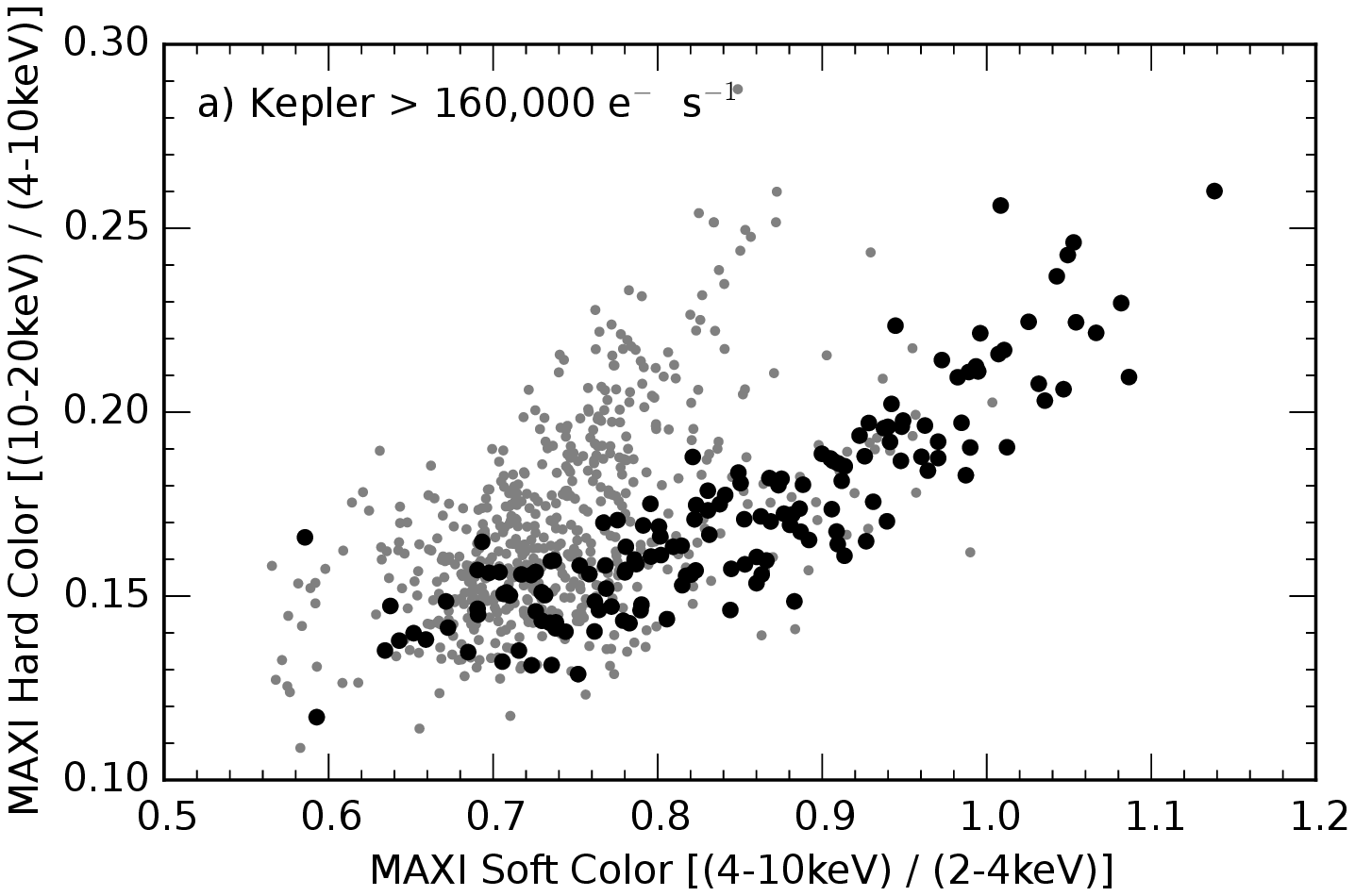}
\epsfig{width=3.5in,file=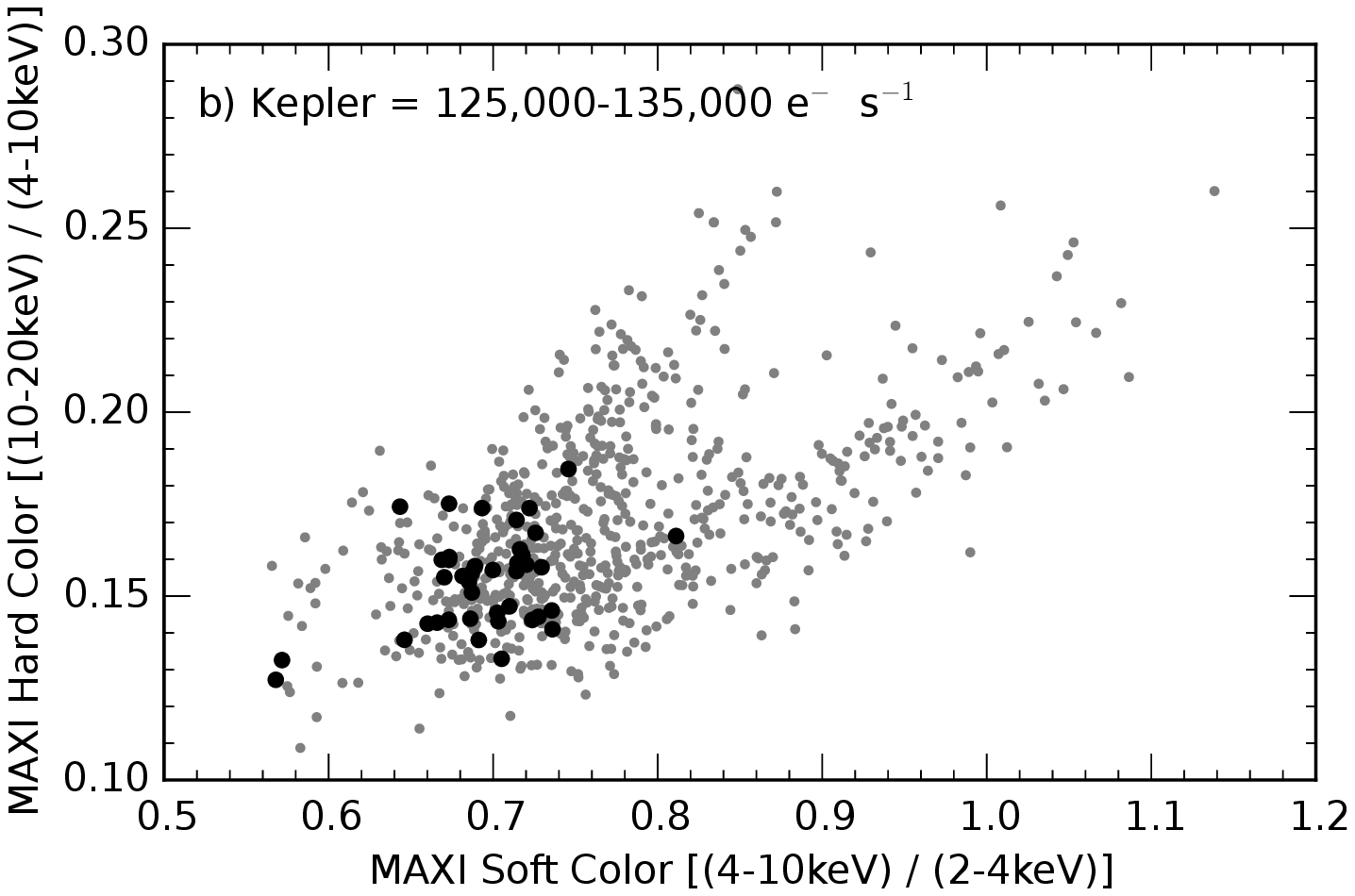}
\epsfig{width=3.5in,file=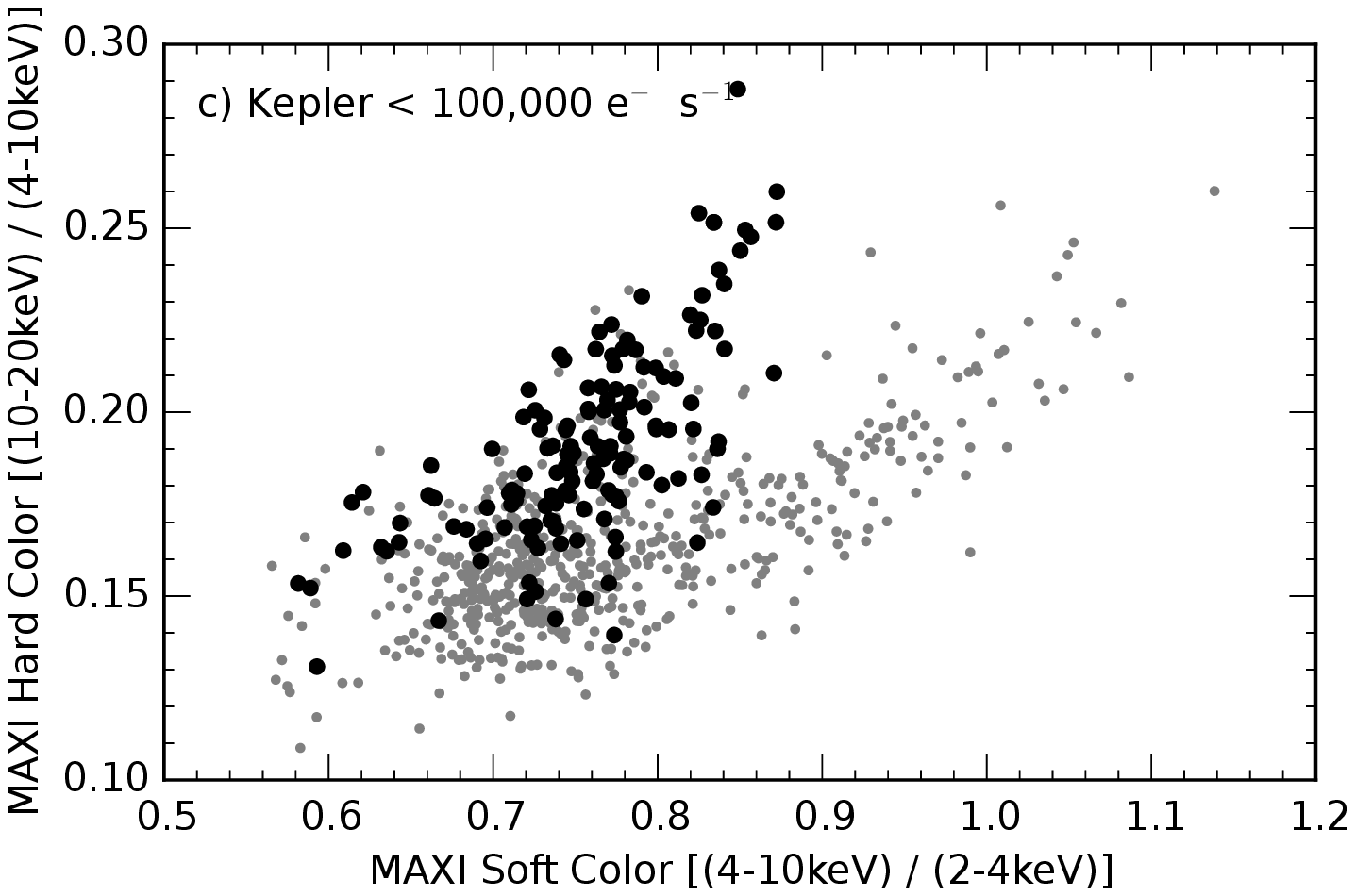}
\caption{The MAXI Z diagram with all data points shown in grey and optically selected points highlighted
  with larger solid black circles as a function of optical
  brightness. This shows that the optical high state corresponds to
  the flaring branch and the optical low state corresponds to the
  normal branch, while optical transitions occur when Sco~X-1 is at
  the soft apex.}
\label{ZFig}
\end{figure}

We relate optical fluxes to X-ray data in a different way in
Figure~\ref{ZFig} where we show Z diagrams derived from MAXI data.  We
highlight subsets of the data by the simultaneous optical flux.  We
see that the lowest optical fluxes, below the fainter of the two peaks
of the optical flux histogram (Figure~\ref{HistFig}), appear to
correspond to the NB, with the upper NB almost always corresponding to
very low optical fluxes.  It is possible that these very low optical
fluxes actually represent the HB, but this cannot be clearly
identified in the MAXI Z diagram. The intermediate flux points between
the peaks of the histogram correspond to the NB-FB transition region
(the soft apex).  The highest flux points correspond to the FB, and
when the source is highest on the FB the optical flux is usually very
high.  This suggests that the optical flux increases monotonically as
Sco~X-1 moves from the hard apex at the top of the NB, down the NB to
the soft apex, and then up the FB.  We will discuss the implications
of this in Section~\ref{DiscussionSection}.

\begin{figure*}
\epsfig{width=3.4in,file=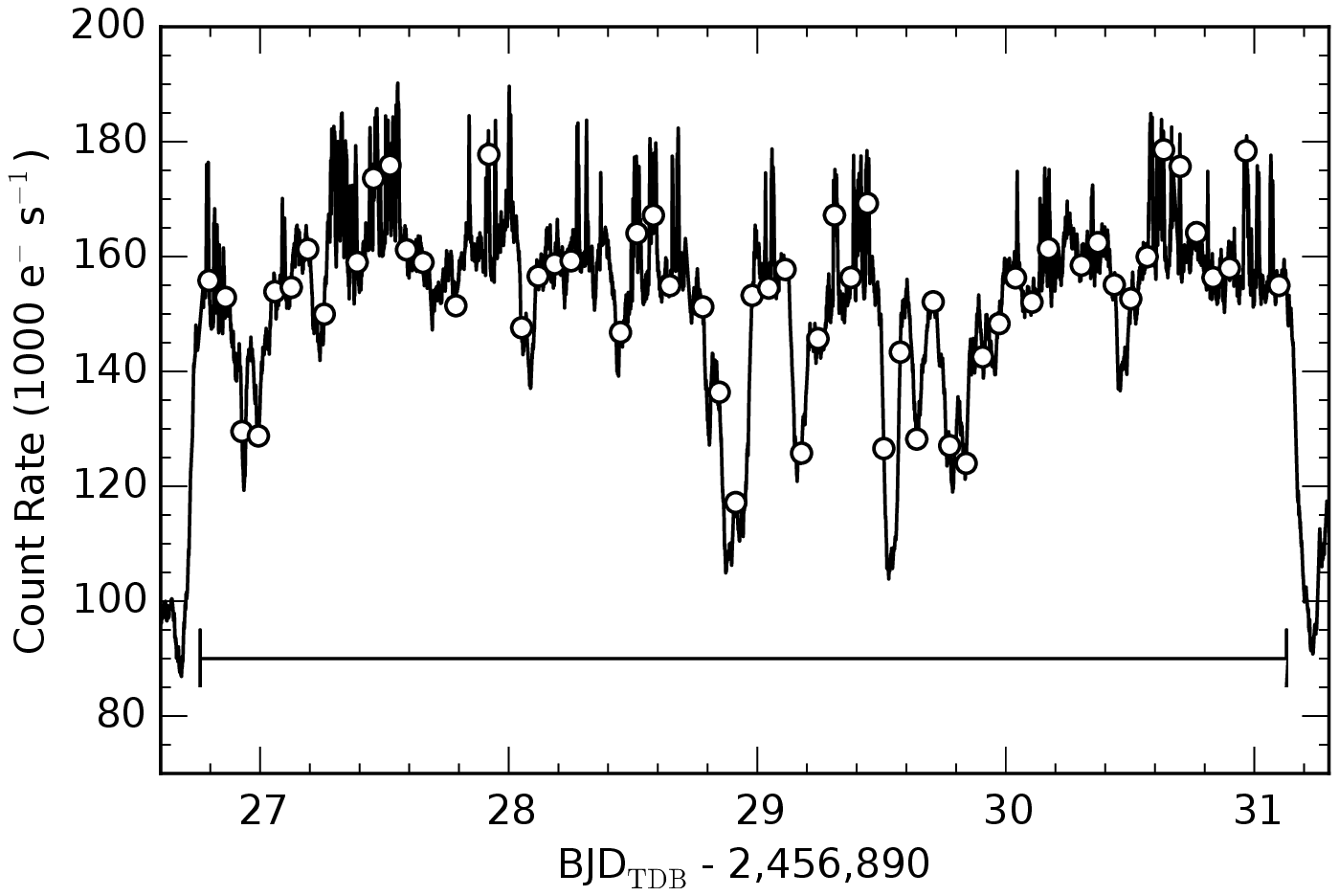}
\epsfig{width=3.4in,file=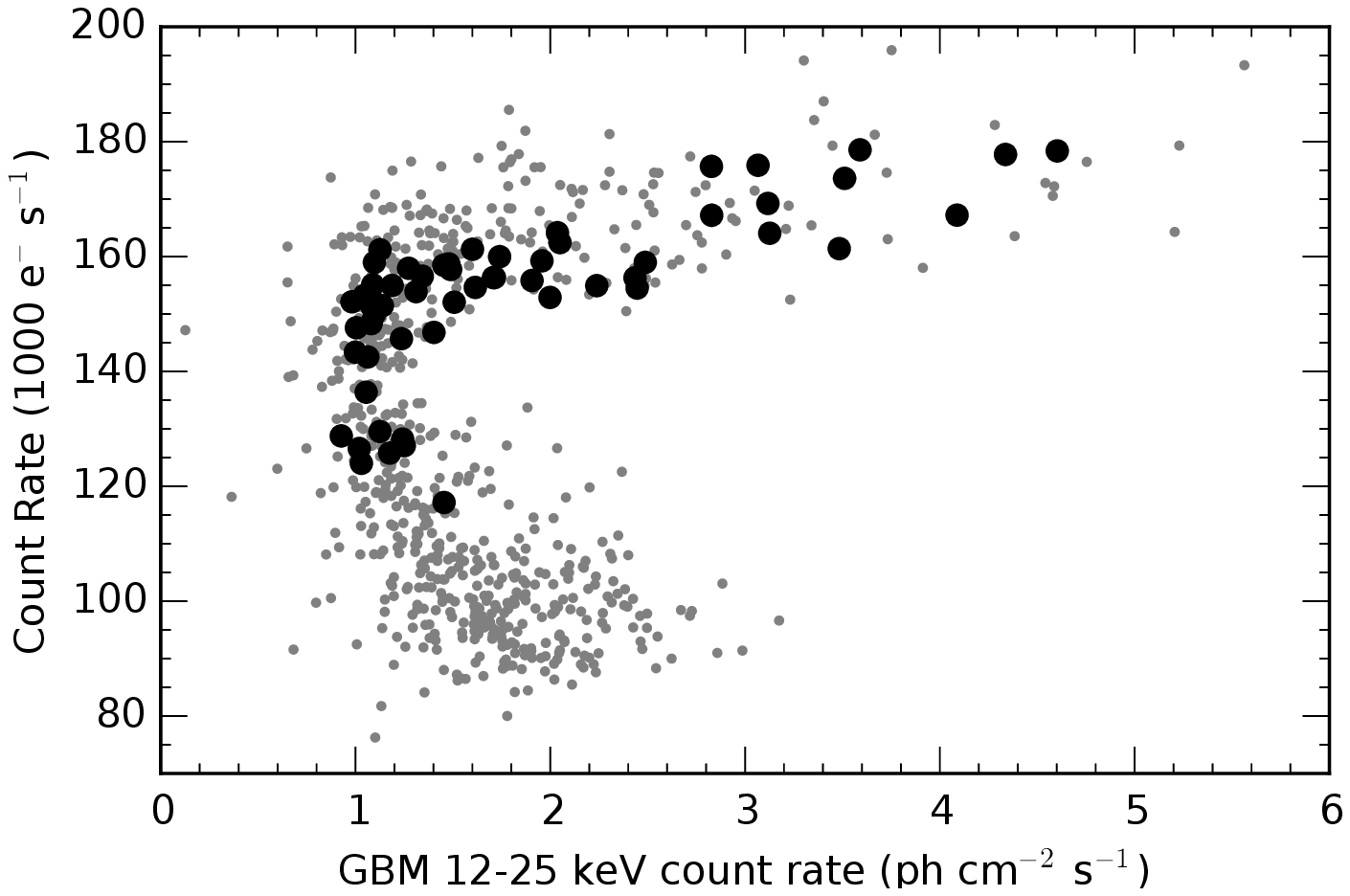}
\caption{Left: Example segment of the {\it Kepler} light curve when Sco~X-1
  was mainly in the high state with a few dips towards the
  low state.  The horizontal bar indicates the range analyzed and the
  open circles indicate the times of GBM occultation and the
  {\it Kepler} flux at that time.  Right: {\it Kepler} vs. GBM flux plot as in
  Figure~\ref{IntensityFig}.  All simultaneous points are shown in
  grey, while GBM observations from the left hand
  panel are highlighted as larger black circles. Several patterns can
  be noted. A very small range of optical fluxes occurs on the
  plateau, during which the X-ray flux varies by about a factor of
  two. During optical flares, the X-ray flux also increases
  substantially. During optical dips, the X-ray flux stays constant or
slightly increases.}
\label{SampleHiFig}
\end{figure*}

Finally, we return to the GBM vs.\ {\it Kepler} flux plot and compare
directly with the {\it Kepler} light curve.  In
Figure~\ref{SampleHiFig} we show a segment of the light curve when the
source was predominantly in the high state, with a lot of flares and
some dips.  GBM times of observation are overlaid at the simultaneous
{\it Kepler} flux.  We also show the X-ray/optical flux plot with the
corresponding points highlighted.  The flares and dips have quite
distinct effects.  During the flares, the X-ray flux increases
strongly, loosely correlated with a moderate optical increase.  During
the dips the optical flux drops substantially, with no change in
X-rays or a moderate increase in X-ray flux.  These changes appear to
correspond to failed transitions to the low state (i.e.  the NB).

\begin{figure*}
\epsfig{width=3.4in,file=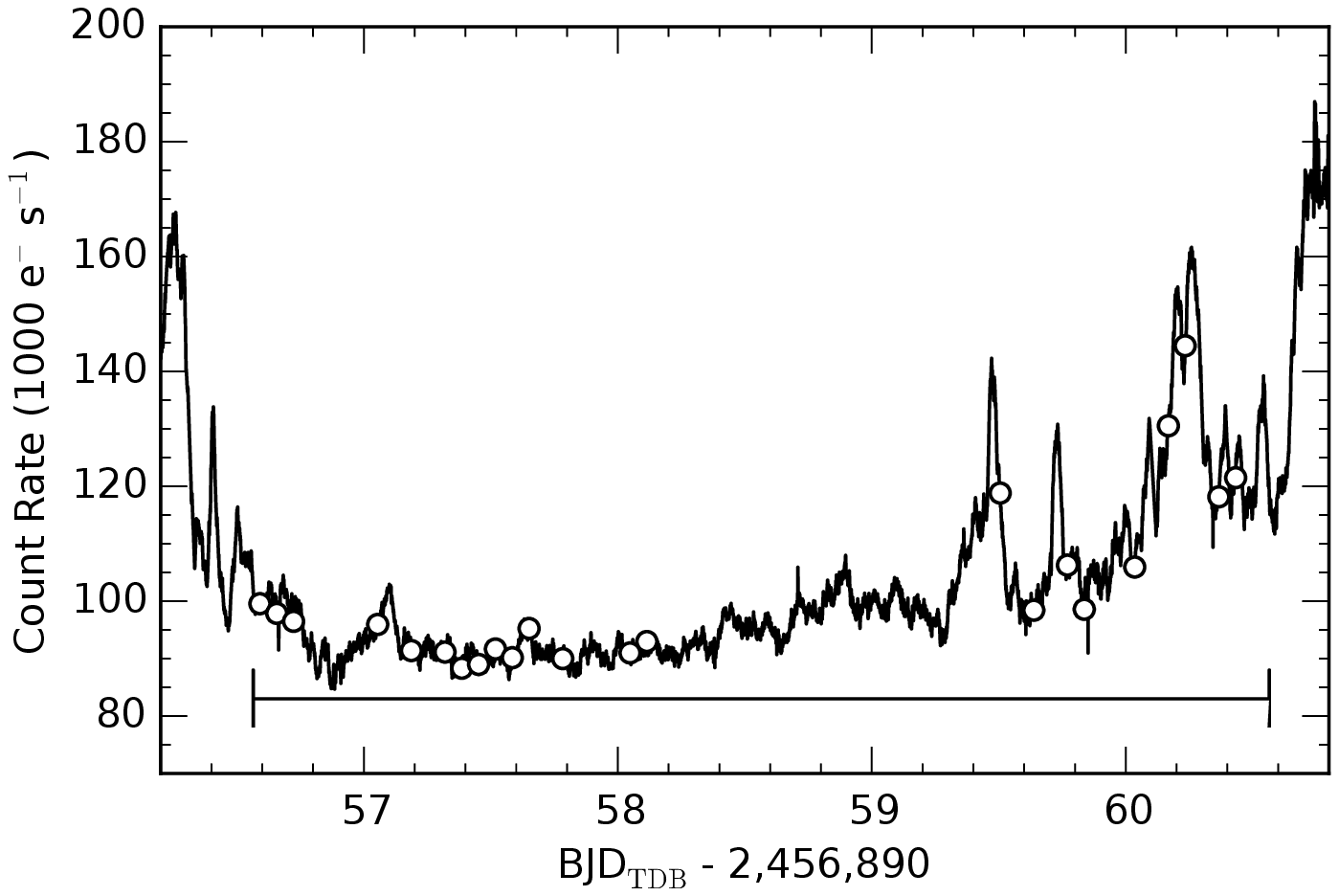}
\epsfig{width=3.4in,file=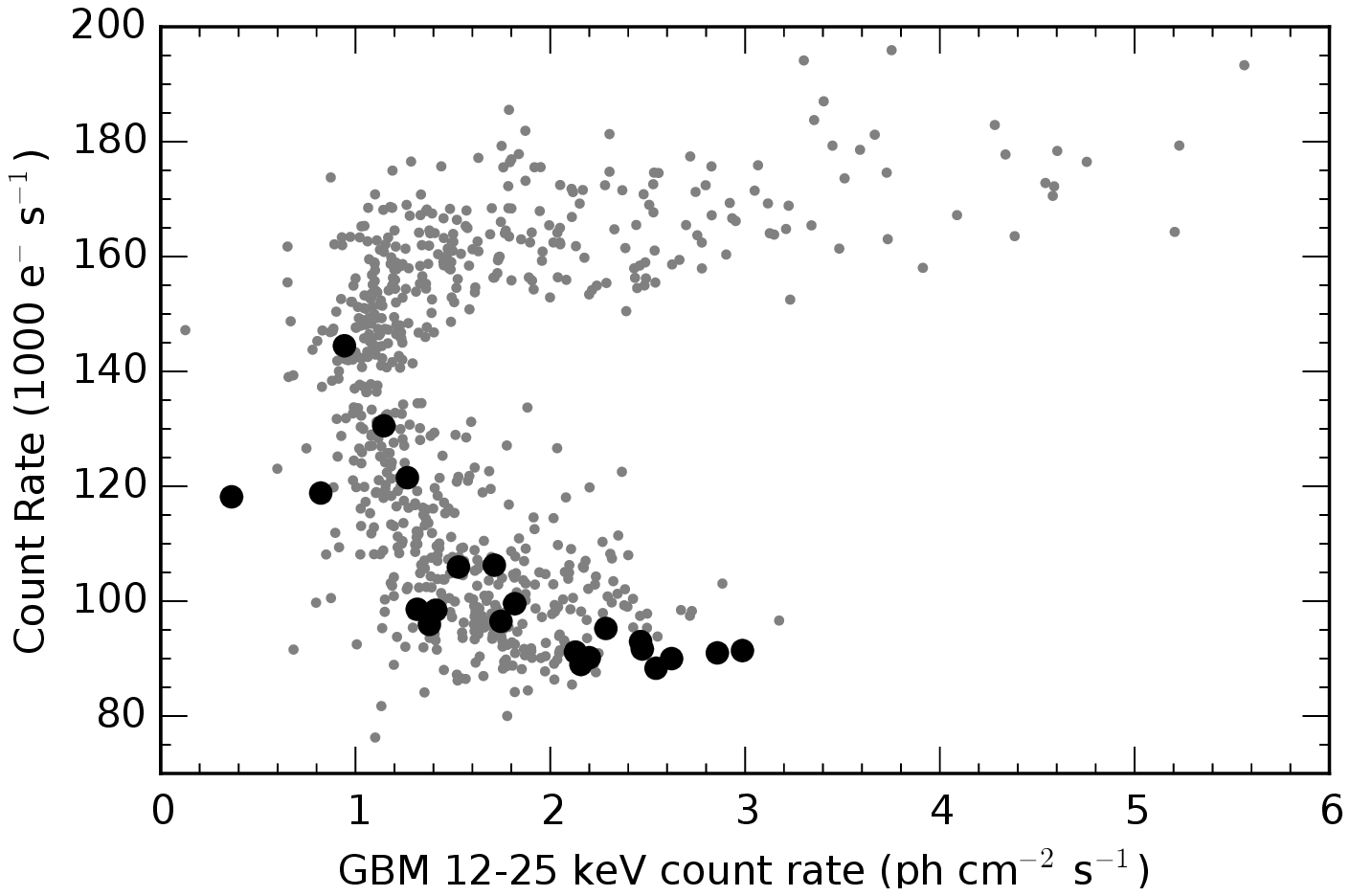}
\caption{Left: Example segment of the {\it Kepler} light curve when Sco~X-1
  was mainly in the low state with a few excursions towards
  the high state.  Right: {\it Kepler} vs. GBM flux plot as in
  Figure~\ref{IntensityFig}. Notation in both plots is as in
  Figure~\ref{SampleHiFig}.}
\label{SampleLoFig}
\end{figure*}

We show a comparable example in the low state in
Figure~\ref{SampleLoFig}.  During this period the source was optically
quite quiescent at the beginning, but there were several failed
transitions to the high state towards the end.  During the quiescent
period at the lowest optical fluxes there are quite large variations
in the hard X-ray flux.  There is little corresponding optical
activity and if anything an anti-correlation.  The anti-correlation is
more pronounced as the source moves towards the high state at the end
of the segment and the hard X-ray flux decreases.

\section{Time delays}
\label{CCFSection}

\begin{figure*}
\epsfig{width=2.3in,file=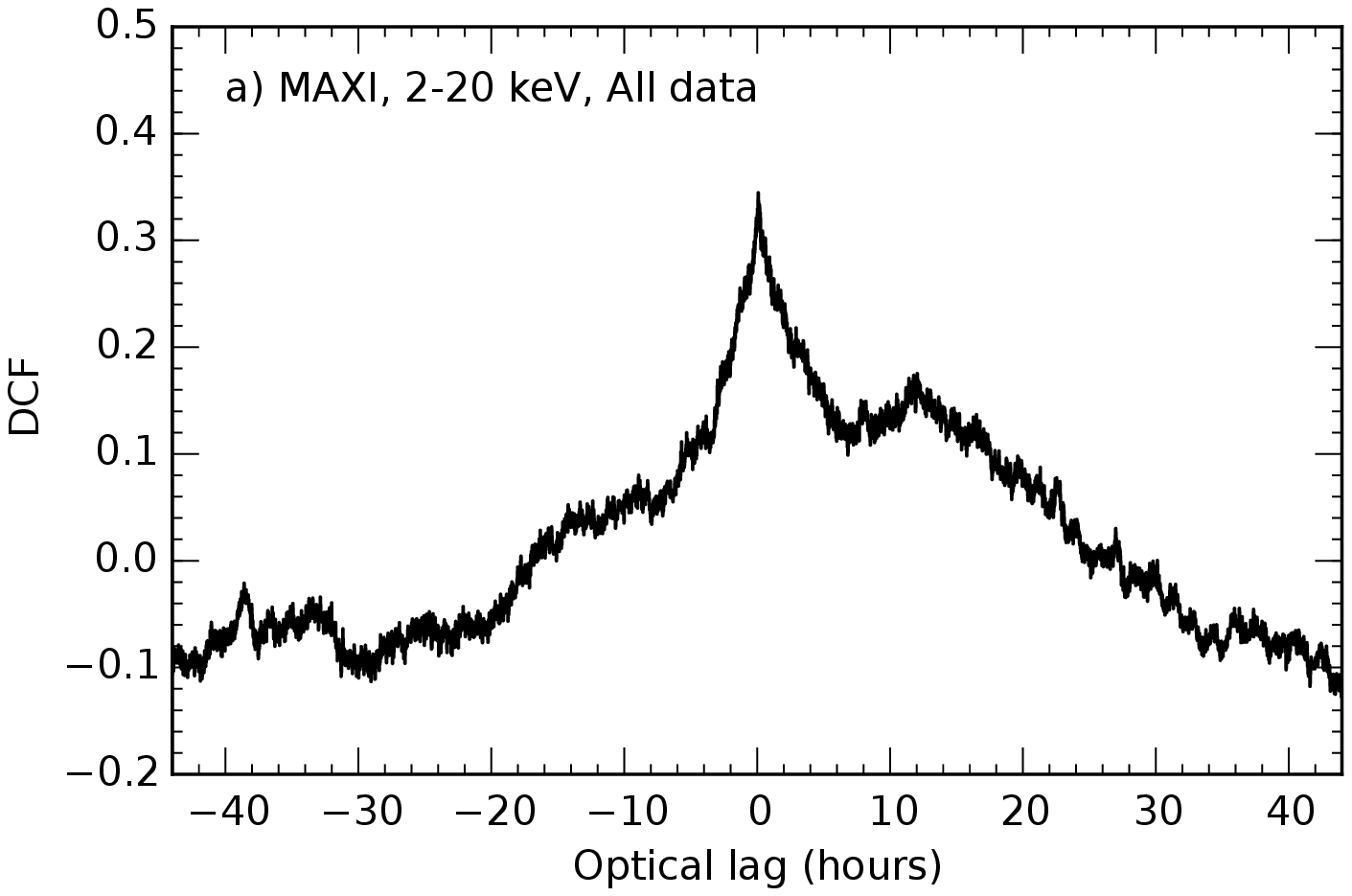}
\epsfig{width=2.3in,file=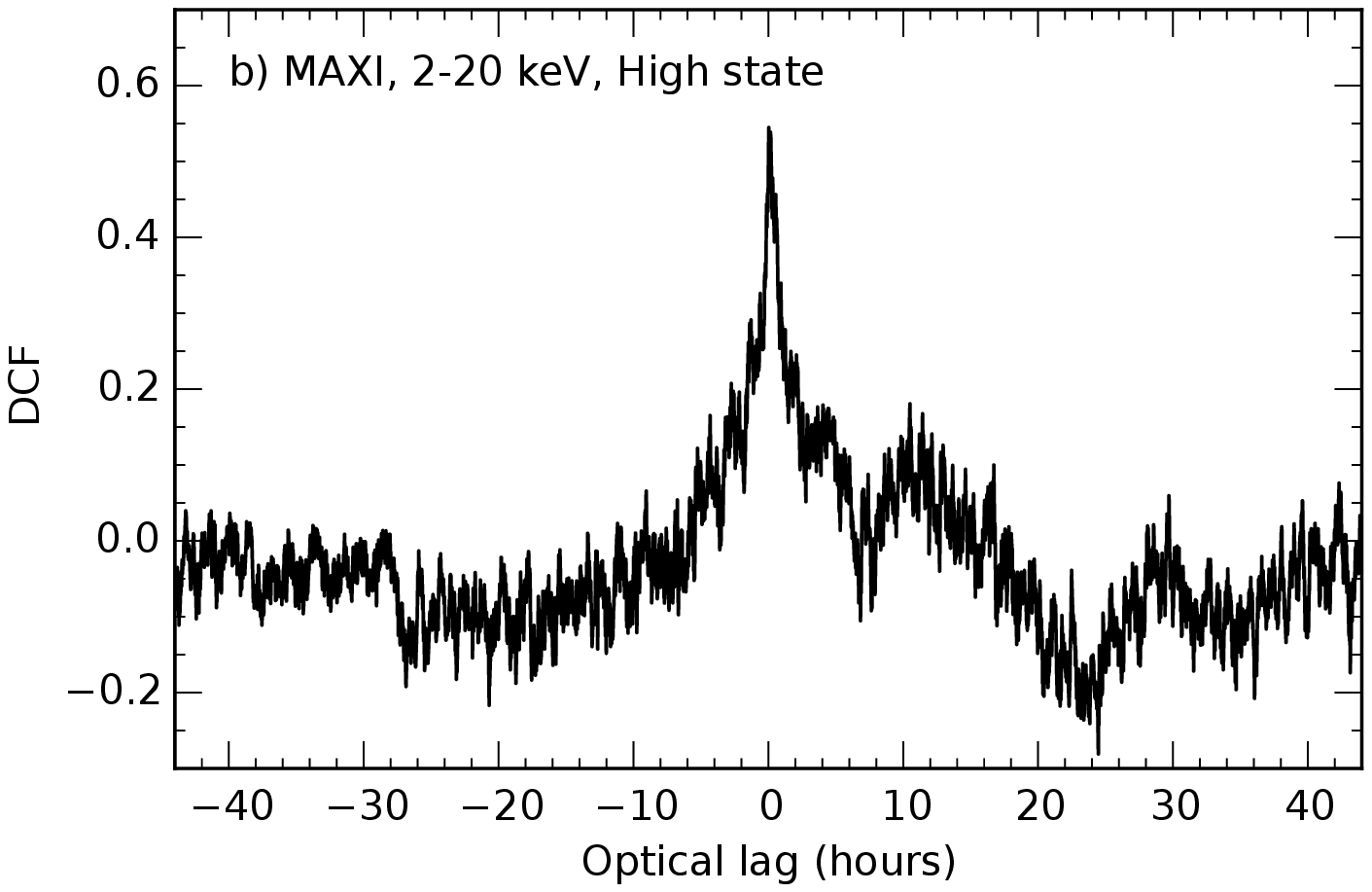}
\epsfig{width=2.3in,file=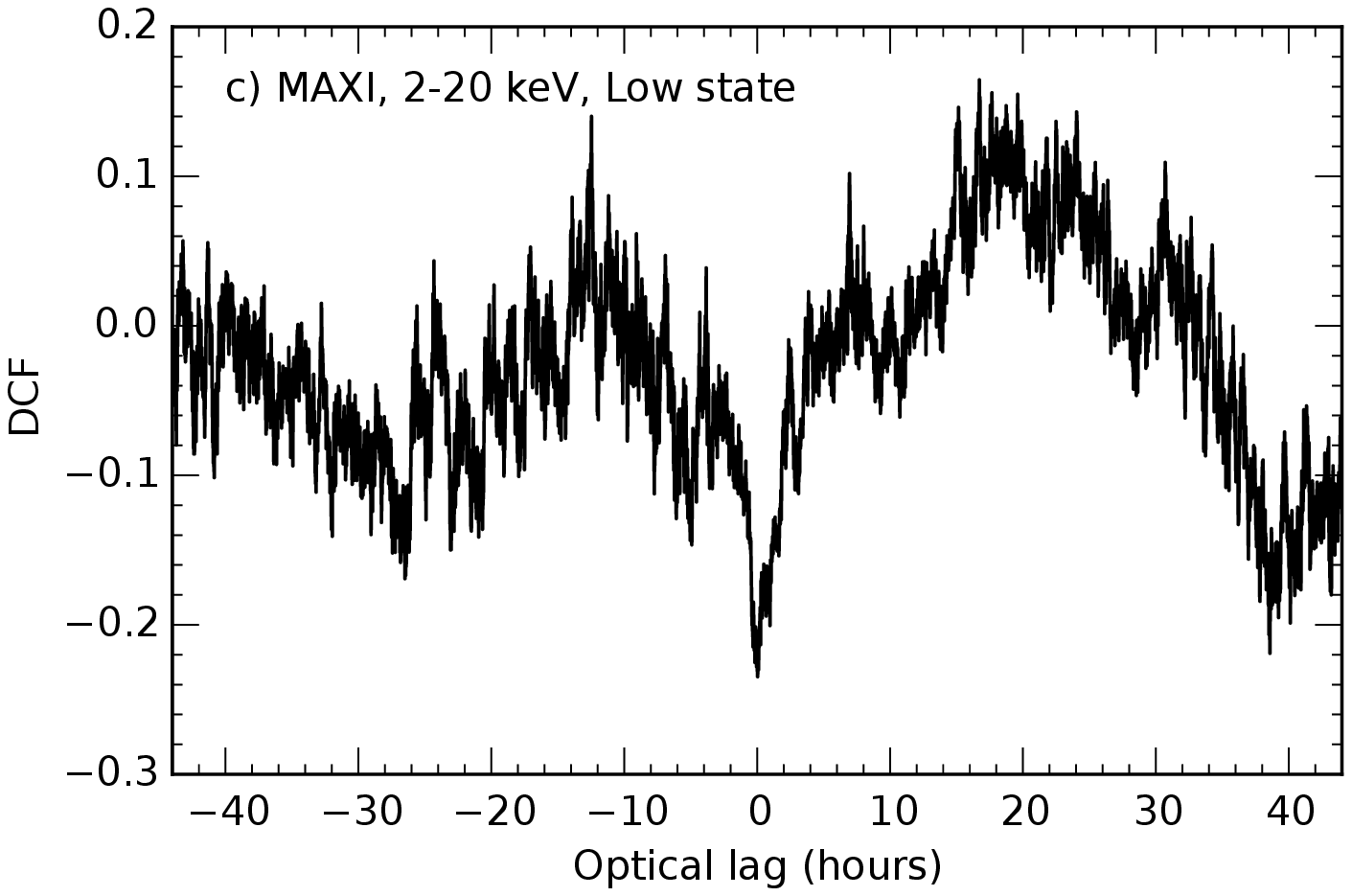}
\epsfig{width=2.3in,file=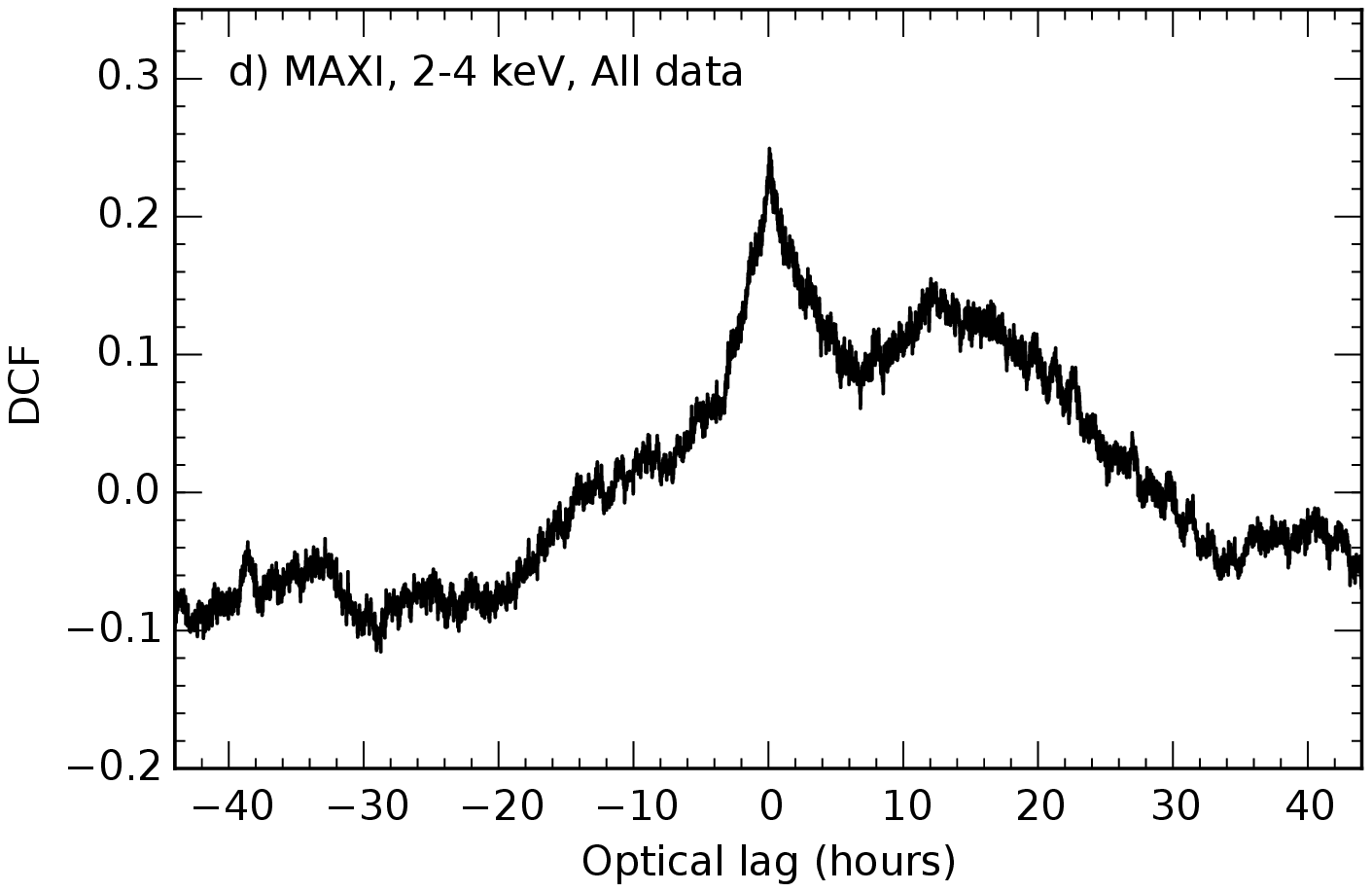}
\epsfig{width=2.3in,file=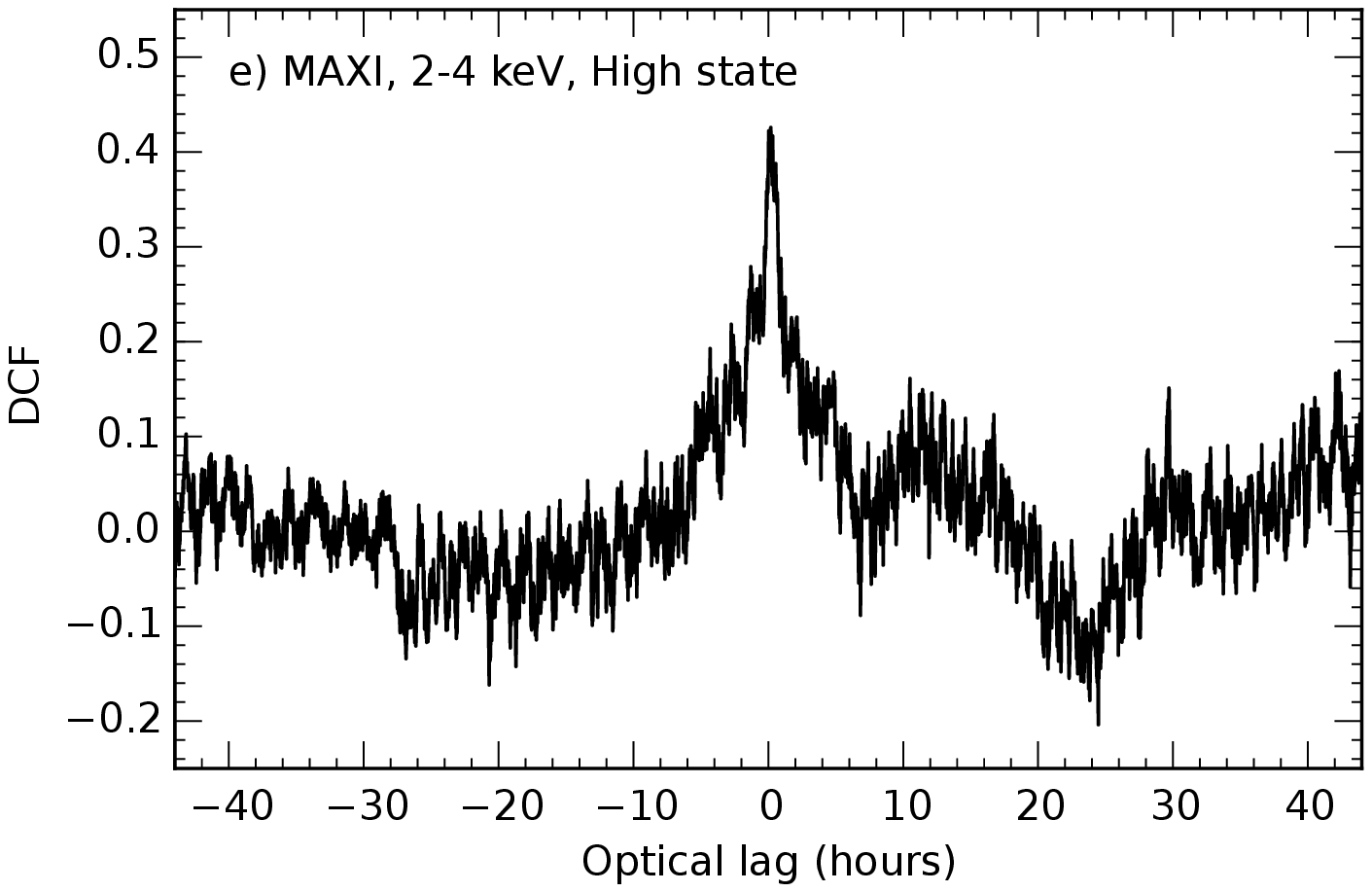}
\epsfig{width=2.3in,file=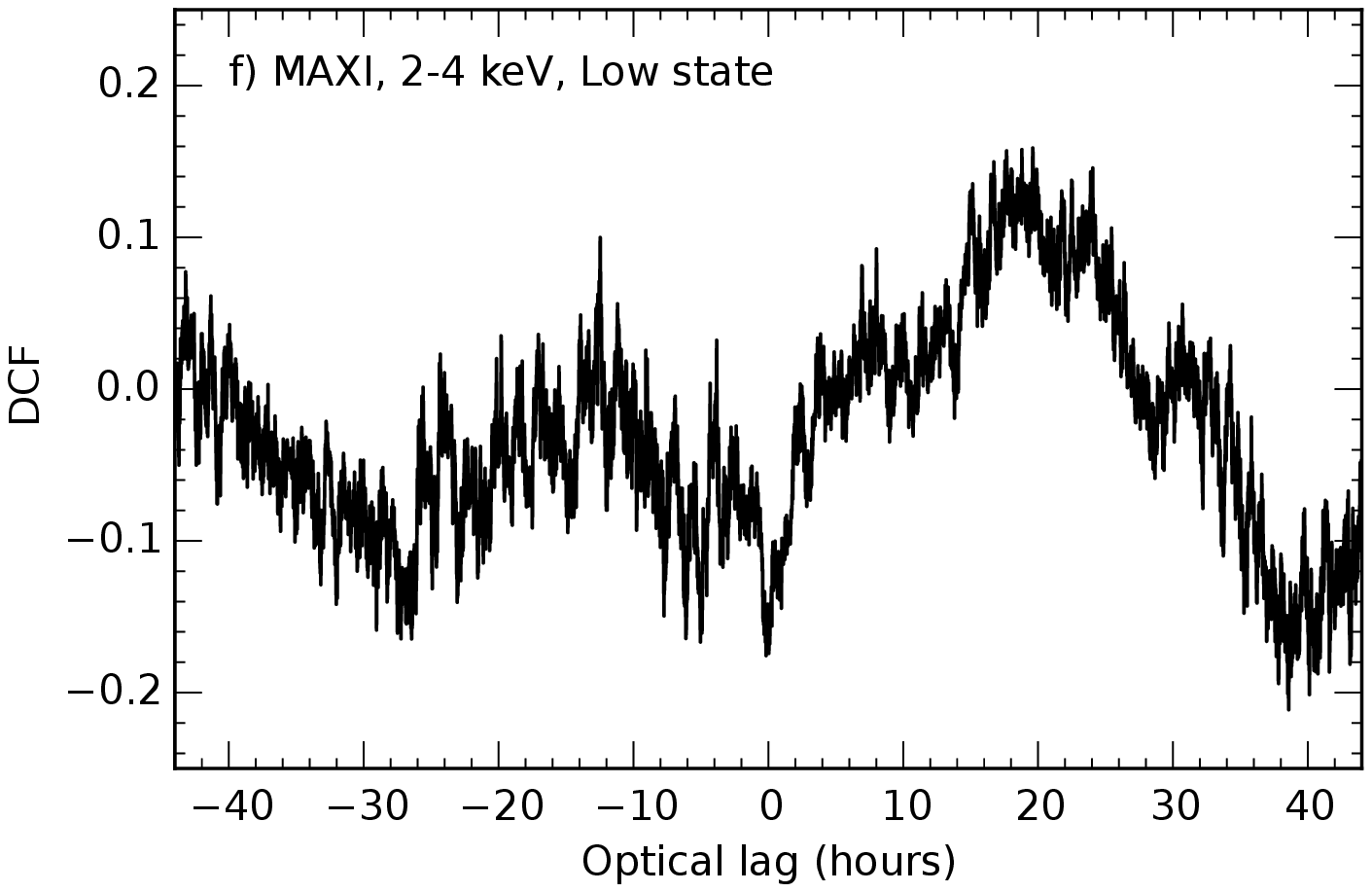}
\epsfig{width=2.3in,file=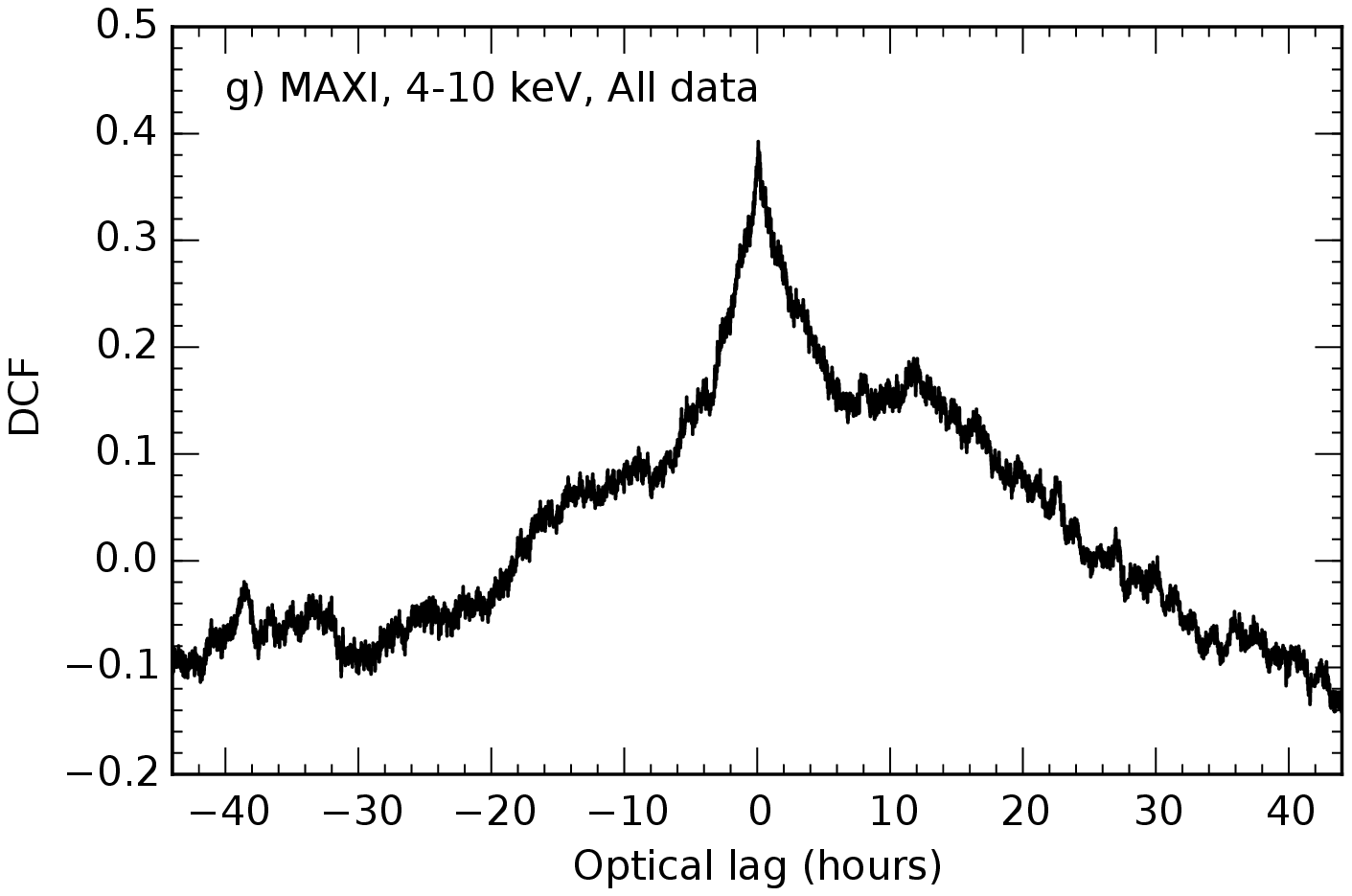}
\epsfig{width=2.3in,file=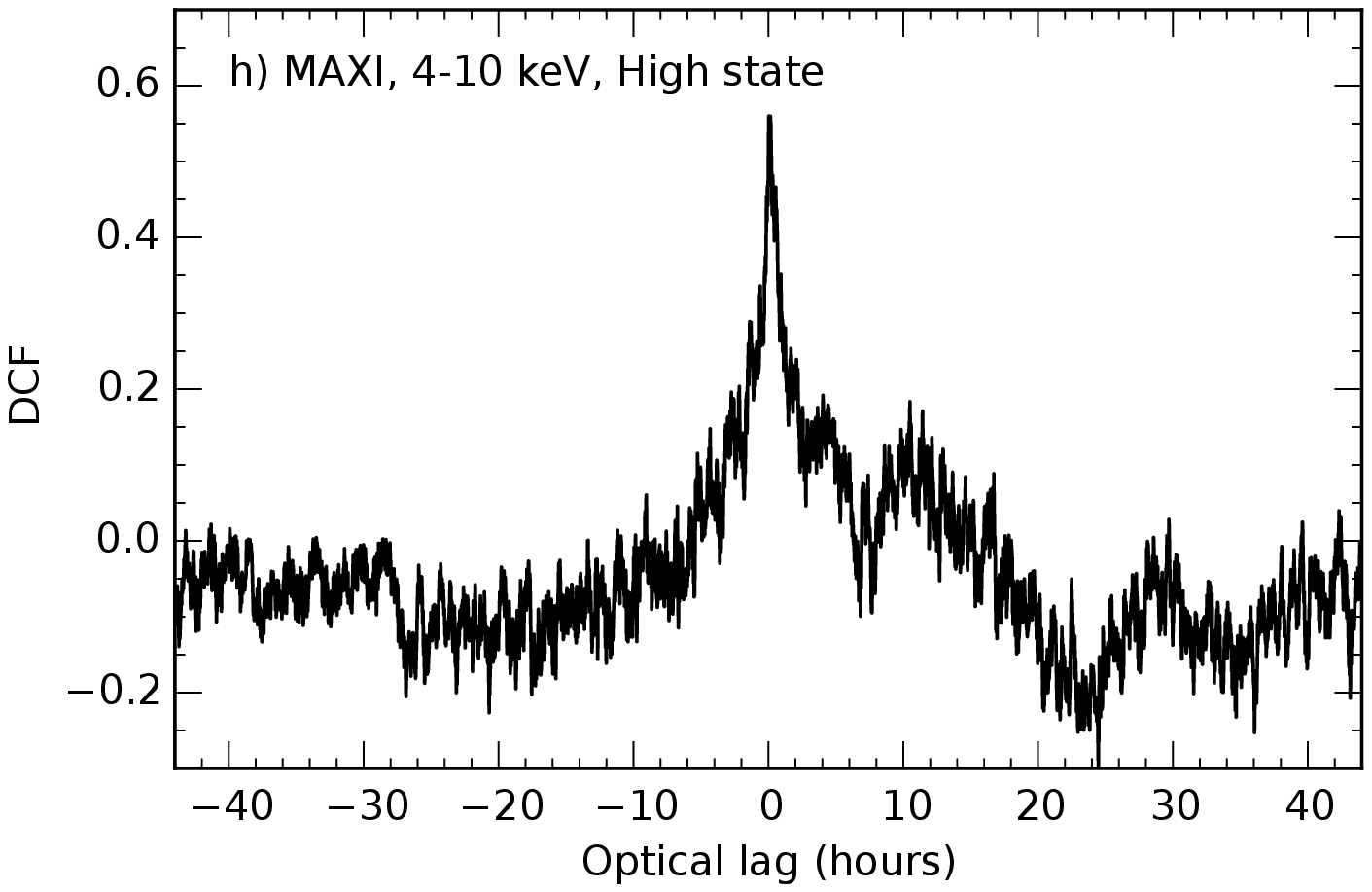}
\epsfig{width=2.3in,file=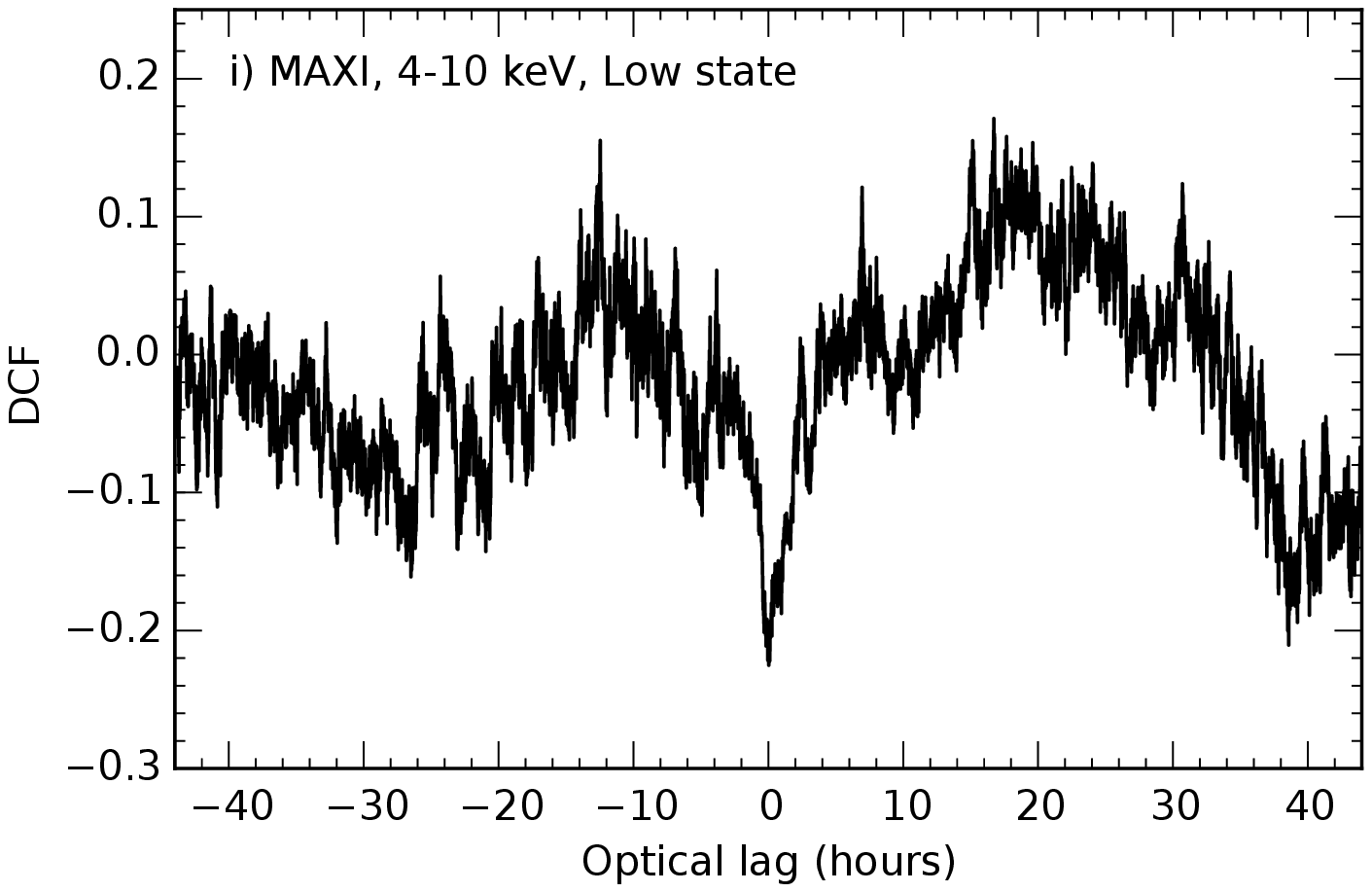}
\epsfig{width=2.3in,file=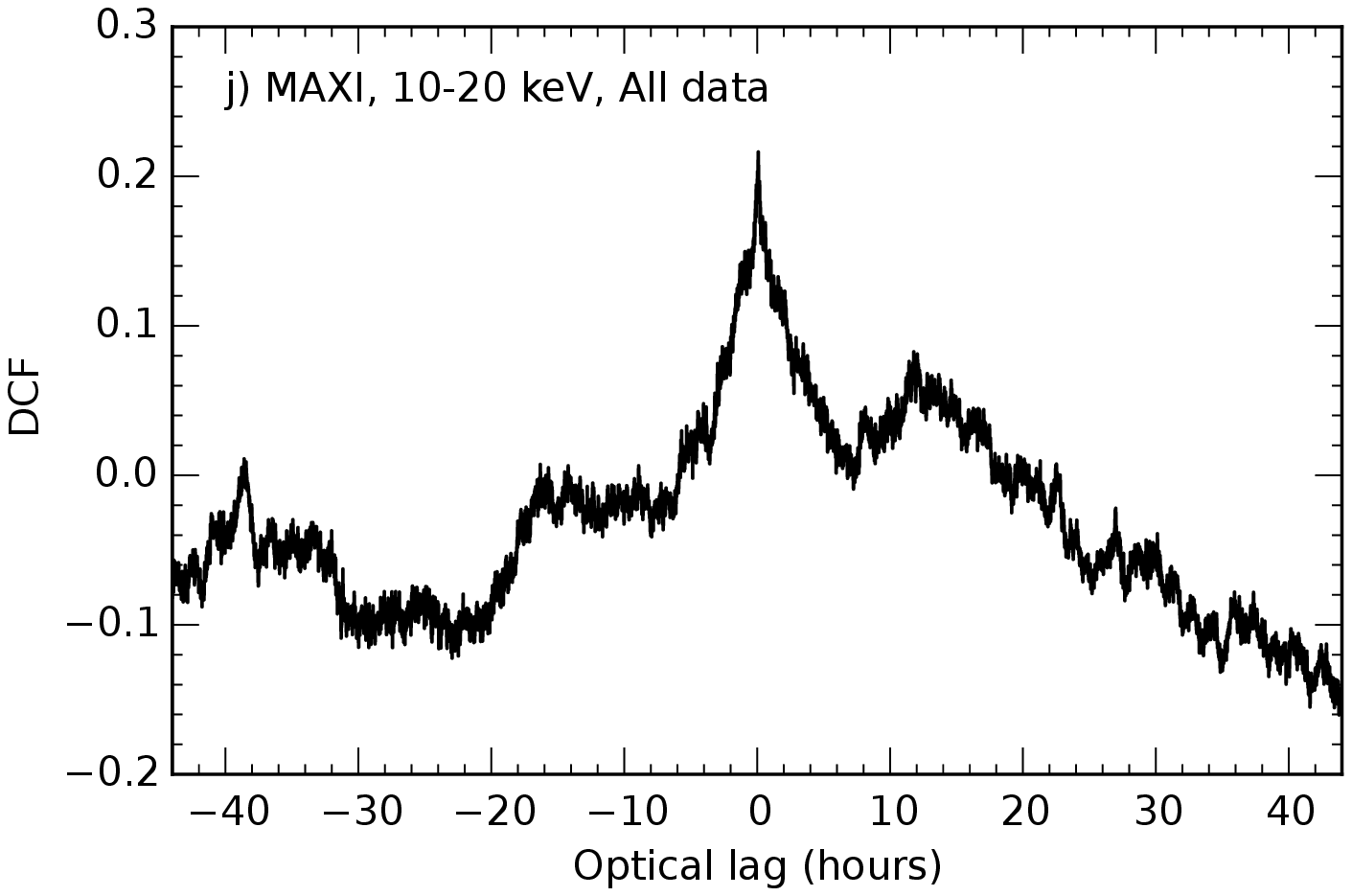}
\epsfig{width=2.3in,file=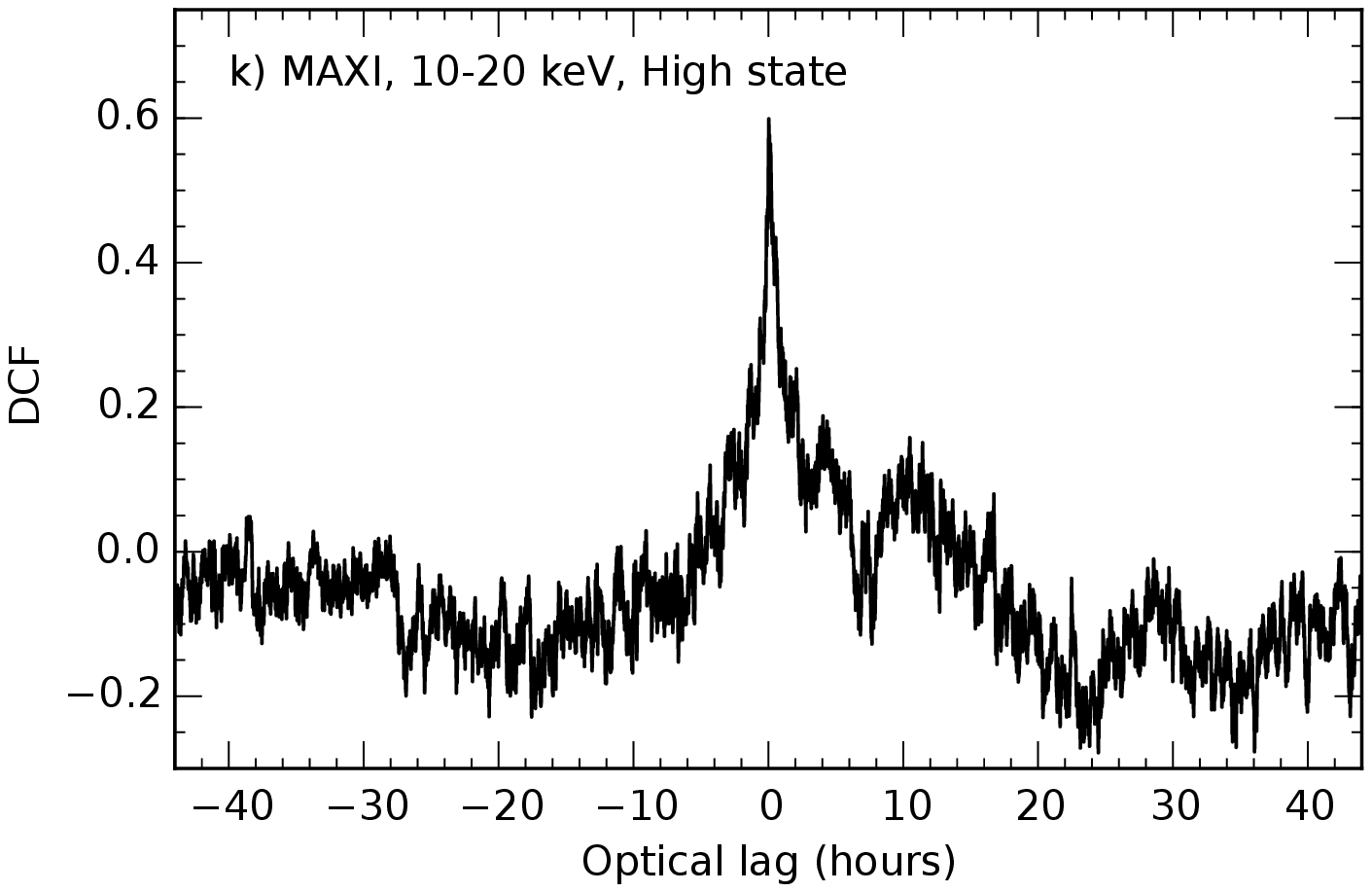}
\epsfig{width=2.3in,file=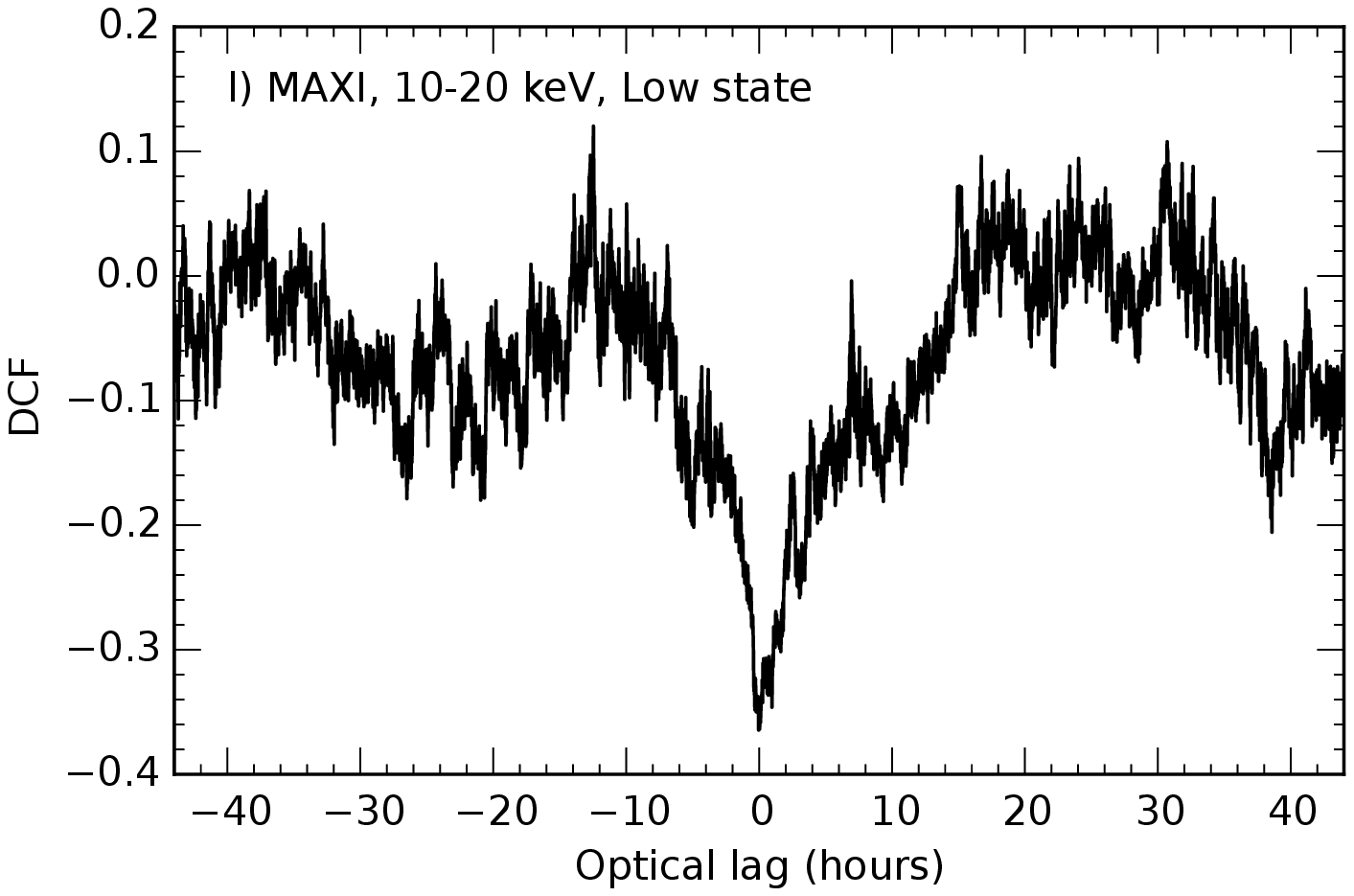}
\epsfig{width=2.3in,file=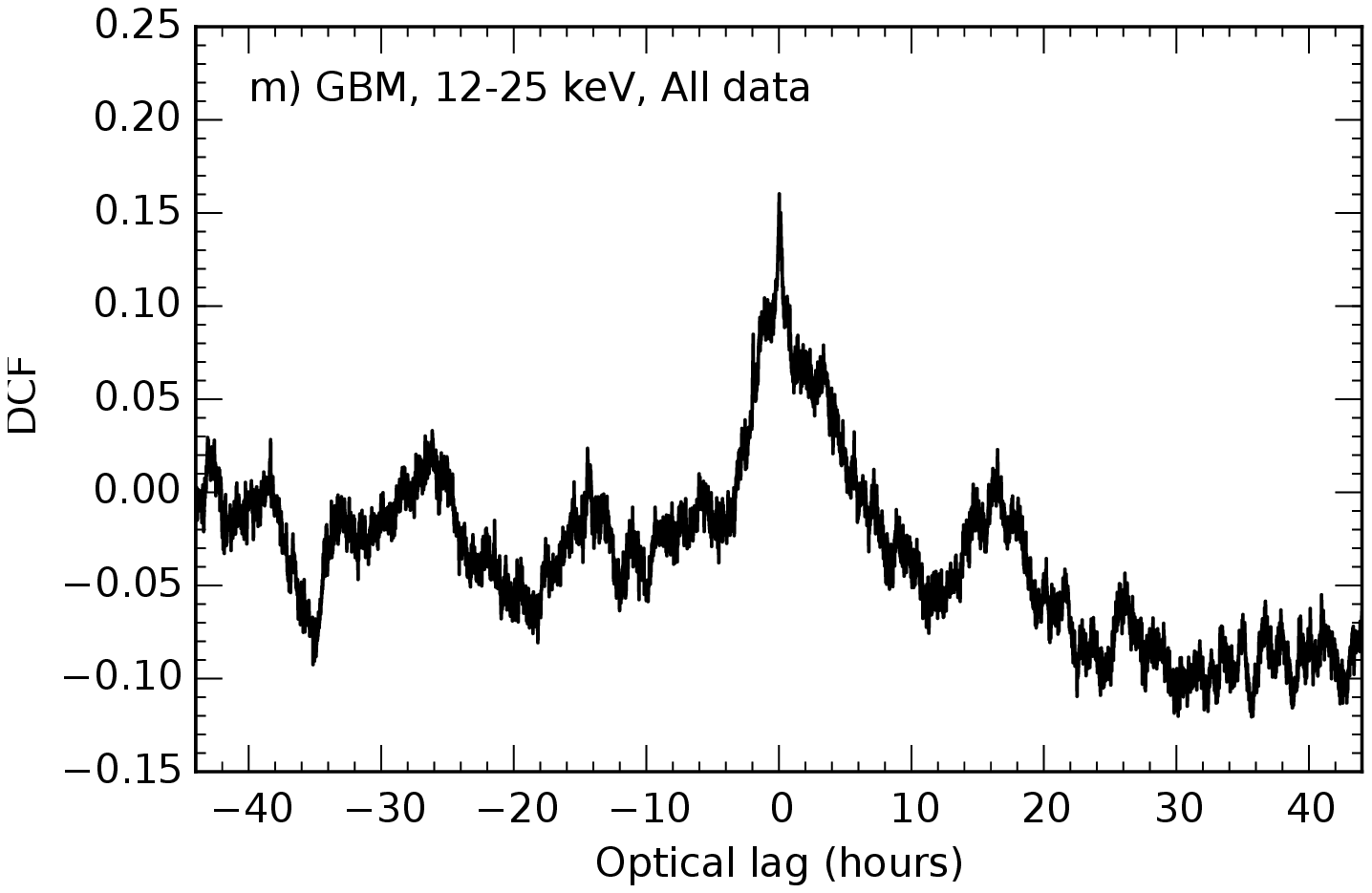}
\epsfig{width=2.3in,file=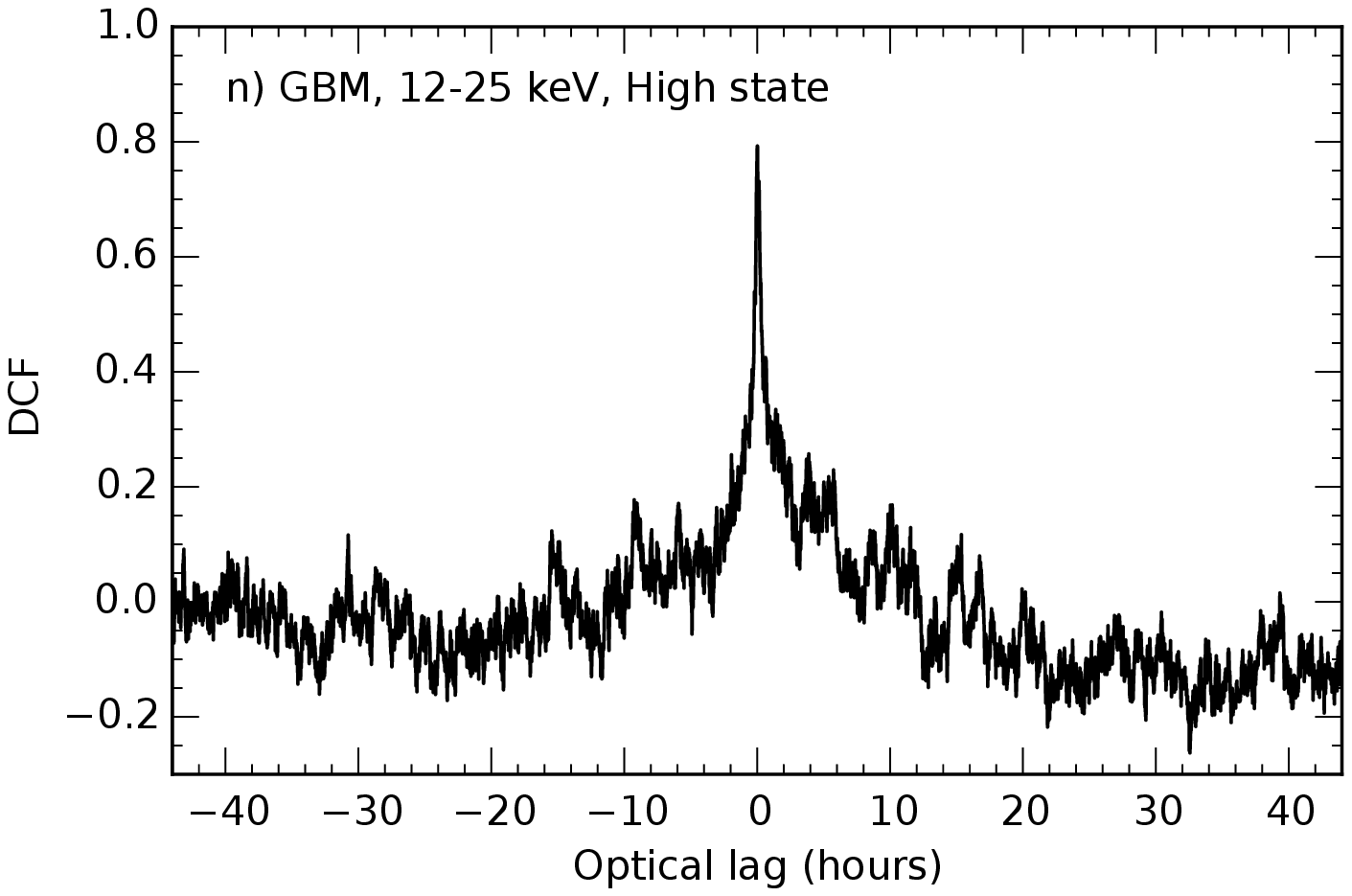}
\epsfig{width=2.3in,file=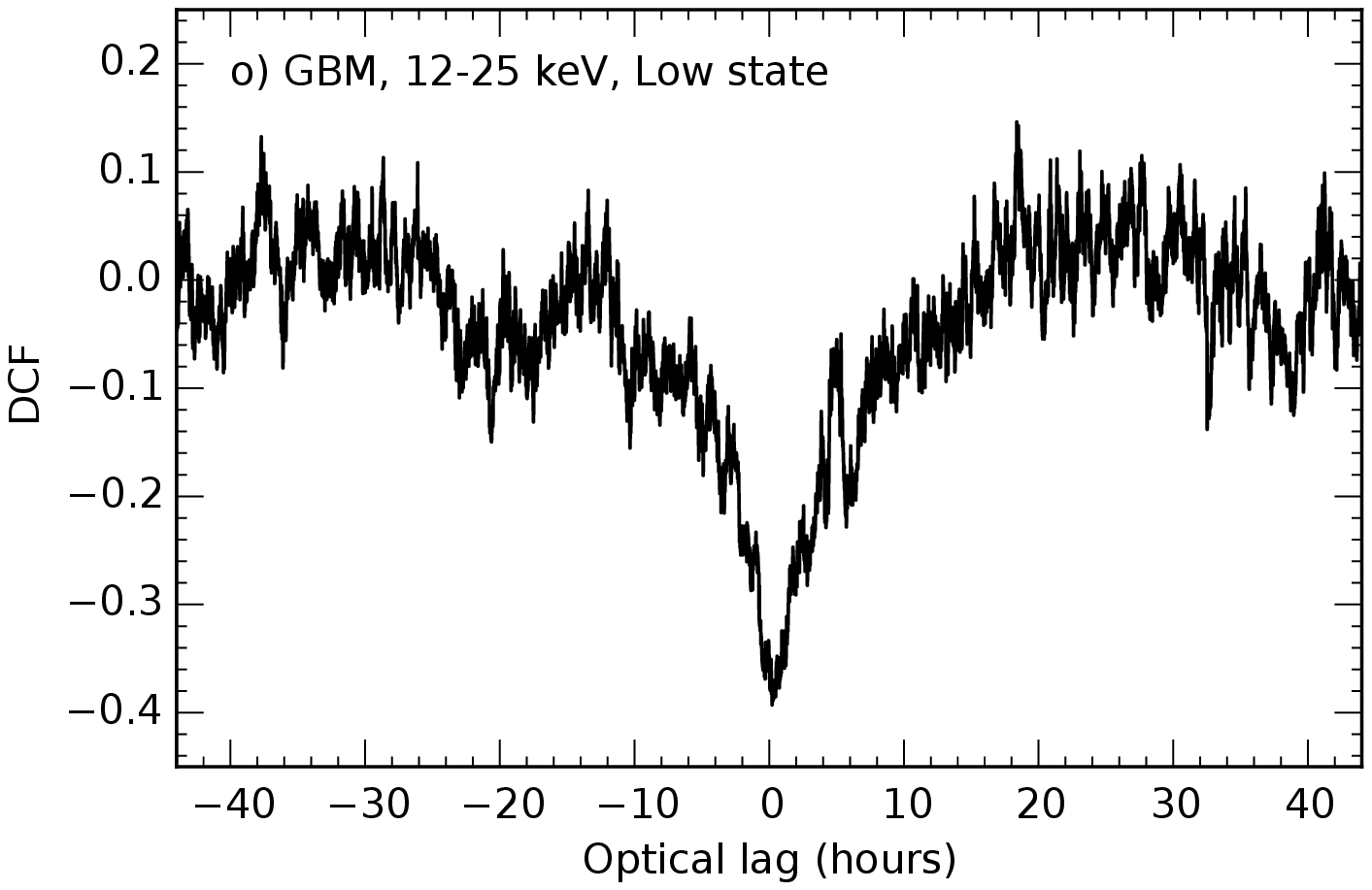}
\caption{DCFs between MAXI or GBM and {\it Kepler} light curves.  The
  left hand panels are for the full data set, the middle panels retain
  only {\it Kepler} points from the high state, and the right hand
  panels retain only {\it Kepler} points from the low state.  From top
  to bottom, we show correlations with the full {\it MAXI} data set,
  the {\it MAXI} low energy data set, the {\it MAXI} medium energy
  data set, the {\it MAXI} high energy data set, and the full {\it GBM}
  data set. These figures clearly show distinct high and low state
  behavior. A sharp positive correlation is always seen in the high
  state data. A broader anti-correlation is always present in the low
  state data, although most prominent when correlating with higher
  energy X-rays.}
\label{CCFFig}
\end{figure*}

Being able to match GBM and MAXI data points to single {\it Kepler}
data points is a modest gain for the flux-flux diagrams shown in the
last section but is a huge gain in constructing cross-correlation
functions (CCFs).  We show the GBM and MAXI vs. optical CCFs in
Figure~\ref{CCFFig}. For MAXI we show both the combined data and
energy dependent data separately to highlight the energy dependence of
the correlations. We use the discrete correlation function (DCF) of
\citet{Edelson:1988a} to pair individual GBM or MAXI and {\it Kepler}
fluxes and then bin the correlations up into 1~min bins to match the
{\it Kepler} resolution.  After removing the optical orbital
modulation we subtract the mean from each light curve and then divide
by the standard deviation before calculating the DCF.  We follow the
recommendations of \citet{White:1994a} and do not weight the
individual data pairs and use the simple standard deviation for
normalization rather than correcting for statistical errors.

Examining the CCFs for the combined data first, these are dominated by
a sharp peak near zero lag, with a broad hump seen clearest in MAXI
combined CCFs for optical lags of about 12~hrs.  This has already been
extensively discussed by \citet{Scaringi:2015a} and attributed to a
thermal time-scale lag. These authors reinforce its significance by
detection of a 12~hr lag in a cepstrum analysis of the {\it Kepler}
light curve.  We reproduce this hump in all the MAXI energy bands, but
not conclusively in the GBM light curve, although a weaker feature is
seen at somewhat longer lags. We were able to reproduce the feature
independently in subsets of the MAXI data, but a bootstrap analysis
using the the stationary bootstrap resampling of both the MAXI and
{\it Kepler} data sets with the average block size set to around 2\,d
in both cases did not reproduce it consistently. Our analysis does not
convincingly confirm or refute this feature. It may be a real, and
possibly transient effect, or it may be a noise peak in the CCF.

Moving on to the CCFs selected by high or low state, we see a number
of clear patterns.  Generally, these are simpler than the complete
CCFs.  In the high state we see a generally positive correlation,
while in the low state we see a dominant anti-correlation.  The
complete CCF will then be the residual remaining after these partially
cancel each other out, so is not of great significance in itself.  We
therefore focus hereafter on the state-dependent CCFs.  These show
general trends with energy, with the higher energies tending to be
simpler and stronger.  The high state is dominated by a single sharp
peak near zero lag. This is present at all energies, but is most
sharply defined in the GBM data.  In the low state, the dominant
feature is an anti-correlation, also near zero lag, but broader than
the high state correlation. This appears to be strongest in higher
energy data and in isolation does not appear significant in the MAXI
2-4~keV data.  The low state also appears to show a broad and
generally positive hump at $\sim20$~hr lags, stronger at lower
energies, but it is unclear if this is a real correlation or CCF
noise. It may contribute to the broad hump in the combined CCFs.

\begin{figure*}
\epsfig{width=3.4in,file=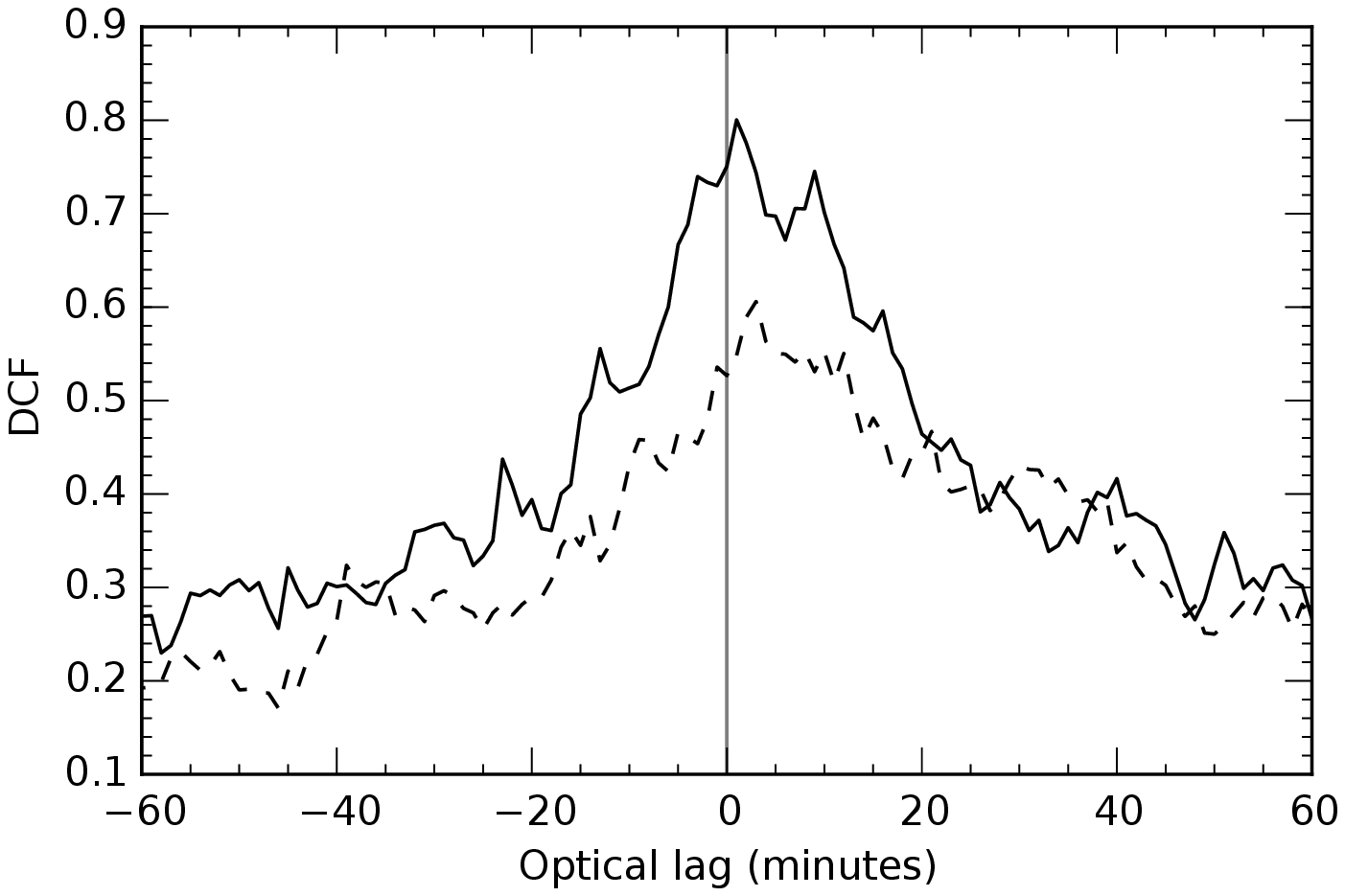}
\epsfig{width=3.4in,file=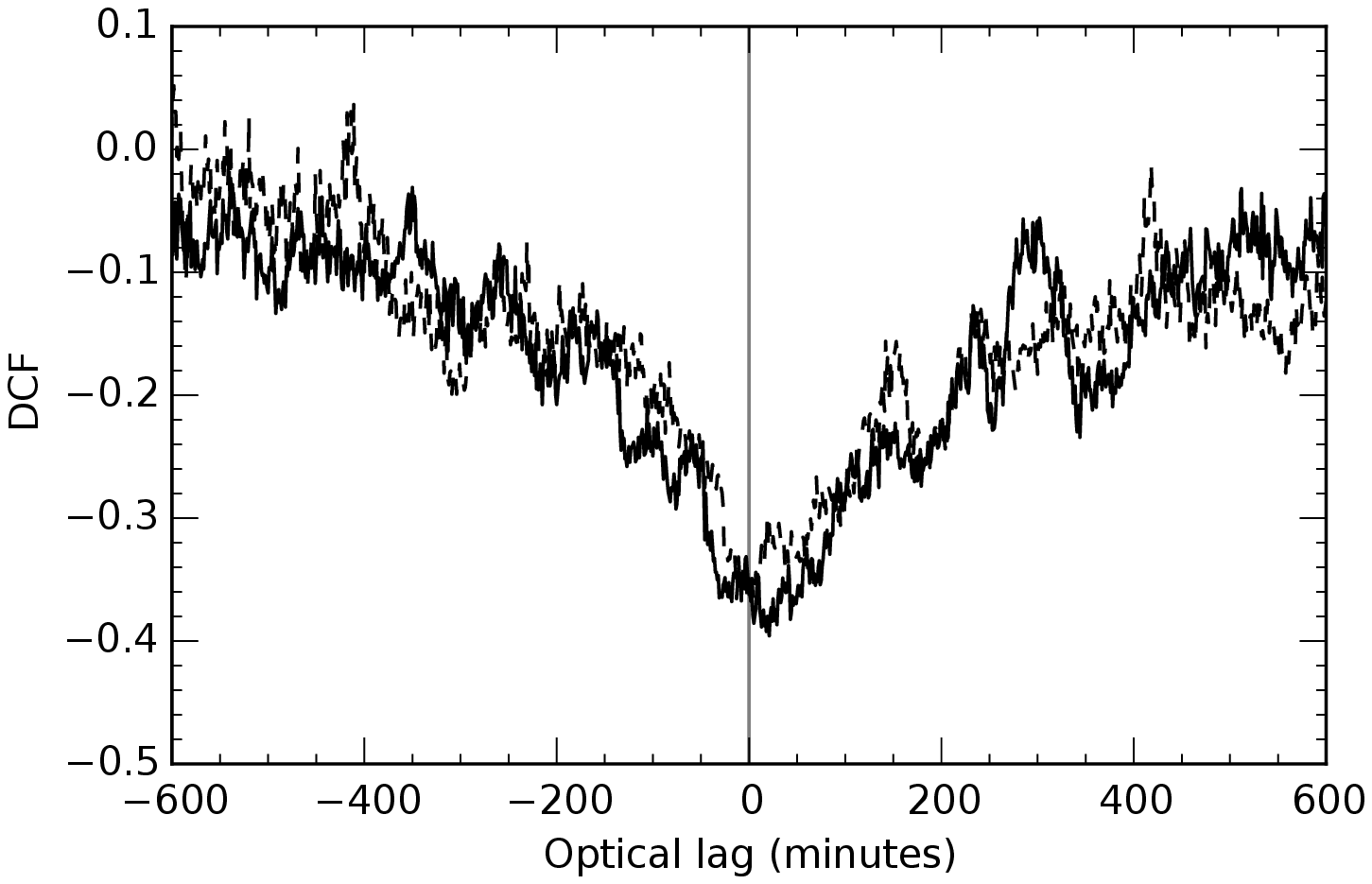}
\caption{Expanded views of the dominant features in the GBM high state
(left) and low state (right) CCF.  Solid lines are GBM data, dashed
lines are MAXI 10--20~keV data.  The vertical line indicates zero
lag. These figures show that the locations and widths of the features
are consistent between the two data sets.}
\label{CCFZoomFig}
\end{figure*}

Since the primary correlations and anti-correlations become best
defined for high energy X-ray data, we will focus on the MAXI
10--20~keV and GBM correlations in more detail.  We show these in the
two states in Figure~\ref{CCFZoomFig}.  We see quite good agreement
between the two satellites based on independent sampling of the
observation period.  In the high state, the positive correlation peaks
near zero lag, with formal peaks at $2.8^{+2.9}_{-2.7}$~min with
respect to GBM data and $6.8^{+4.5}_{-3.9}$~min with respect to MAXI
data.  Uncertainties were estimated using a stationary bootstrap
resampling of both the GBM/MAXI and {\it Kepler} data sets with the
average block size set to around 2\,d in both cases. The measurements
suggest a modest optical lag, but do not individually rule out zero
lag with high confidence. We can quantify this more formally by taking
weighted averages of the lags from pairs of GBM and MAXI bootstrap
lags. We find 4.8~percent of the sample produce an average lag below
zero, so zero lag can be ruled out at 95~percent confidence, but this
still leaves open the possibility that this is a coincidence with a
5\,percent probability.

In the low state, both satellites agree on the presence of a broad
anti-correlation, with FWHM $\sim5$~hrs, much broader than the positive
correlation in the high state (FWHM $\sim25$~min).  The difference in
widths of the features reflects the different characteristic
time-scales; in the high state flares have durations of typically
5--20~min (Section~\ref{FlareSection}), while low state brightenings,
dips, and transitions are much slower. The low state CCF also suggests
that the optical might lag the X-rays.  We measure an optical lag of
$29^{+30}_{-28}$~min with respect to the GBM data and
$37^{+47}_{-43}$~min with respect to the MAXI data.  Errors were
estimated in the same way as for the high state and in this case
11~percent of bootstrap trials yielded a weighted average of the GBM
and MAXI lags below zero. Thus although the data are consistent with
quite a large lag, around 30~min, zero lag cannot be ruled out even at
90~percent confidence.

\section{State dependent X-ray spectroscopy}
\label{XraySpecSection}

\begin{figure}
\epsfig{width=3.5in,file=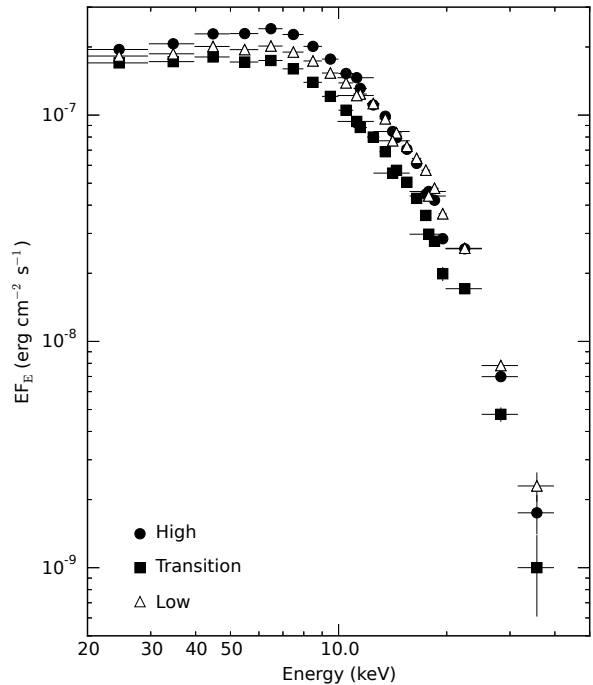}
\caption{MAXI and GBM spectra selected according to state. Open
  symbols indicate MAXI data, closed symbols are GBM data. This figure
demonstrates that it is possible to use optical state classifications
to separate high energy data by state.}
\label{GBMFig}
\end{figure}

Using the high and low states as proxies for the X-ray FB and NB
respectively as described in Section~\ref{FluxDiagramSection}, it is
possible to construct spectra of the two states as shown in
Figure~\ref{GBMFig}.  A third `transition' state spectrum was created
using data that were intermediate between the high and low states.
MAXI data (2--20~keV)and GBM/CSPEC data (12--40~keV) were used to
create the spectra.  The data were separated into the appropriate
states using the {\it Kepler} data to determine which state the source
was in for a given MAXI or GBM observation and then an average
spectrum was constructed for each state and each observatory. Based on
the Z diagram locations associated with optical states, we expect the
high state spectrum to be an average flaring branch spectrum. The low
state spectrum should be an average normal branch spectrum, and the
transition spectrum should be a spectrum at the soft apex.

The three spectra generally show more similarities than differences,
reflecting the modest X-ray changes compared to the quite dramatic
optical variation between states.  All are dominated by a single
quasi-thermal component. This is consistent with findings of other
groups for the energy ranges considered \citep[e.g.][and references
therein]{Church:2012a,Titarchuk:2014a}. For example,
\citet{Titarchuk:2014a} models this quasi-thermal component as a
Comptonized black body and finds it dominates the spectrum in all
states below $\sim40$~keV. It is interpreted as black body emission
from the neutron star surface Comptonized by hot material in the inner
transition layer.  \citet{Church:2012a} also attribute this component
to the neutron star and model the spectrum as a black body with a
cut-off power-law tail, subject to absorption.

The differences we see between the states can readily be associated
with the color changes in the Z diagram. The optical transition state
corresponds to the soft apex of the Z diagram when the source is
transitioning between NB and FB.  This has the softest colors in the Z
diagram, and the lowest X-ray flux, so is the faintest state at all
energies.  As Sco~X-1 moves up the FB (high state), its soft color
increases significantly while the hard color only increases a
little. The intensity also increases. The resulting overall higher
intensity, and harder spectrum below 10~keV compared to the transition
state can readily be seen in Figure~\ref{GBMFig}.  In the NB (low
state), the hard color increases more pronouncedly than the soft
color, with a less pronounced intensity change. As a result it has
softer soft colors than the FB (a flatter low energy spectrum) and
harder hard colors (a slower drop off at high energies), and so
initially is flatter but overtakes the FB at high energies.

The spectral differences we find are entirely consistent with more
comprehensive X-ray spectral analyses \citep[e.g.][and references
therein]{Church:2012a,Titarchuk:2014a}. The key innovation is that we
have shown that state-dependent hard X-ray spectra can be constructed
using optically defined states only. In fact, since the optical
transition seen near the soft apex is quite dramatic, the optical may
allow a cleaner state selection than is possible using X-ray
colour-colour diagrams, though probably still inferior to a full
spectral-timing classification.

\section{Discussion}
\label{DiscussionSection}

Before attempting to tie together the results obtained, let us review
the disc geometry and time-scales.  For Sco~X-1, assuming orbital
period $P=18.9$~hrs, mass ratio $q=0.5$, and neutron star mass
$M_1=1.4$~M$_{\odot}$ \citep{MataSanchez:2015a}, the binary separation
is about $3\times10^{11}$~cm and the disc radius is about
$1.3\times10^{11}$~cm.  Light travel time delays should then be about
10~s to the companion star and a few seconds to the disc.  For typical
disc edge temperatures in LMXBs (10\,000--20\,000~K) and temperature
distributions (between $T\propto R^{-3/4}$ and $T\propto R^{-1/2}$)
the $V$ band flux weighted average disc radius should be about half of
the total disc radius.  This is about $6\times10^{10}$~cm and should
be representative of where typical optical photons originate (if they
arise from thermal disc emission).  At this radius we expect a
dynamical time-scale, $\tau_{\rm d}$ of about 20~min
\citep*{Frank:2002a}.  The thermal time-scale, $\tau_{\rm th}$ should
be $\tau_{\rm d}/\alpha$, where $\alpha$ is the viscosity parameter
\citep{Shakura:1973a}.  The $\alpha$ parameter can be as large as
unity, although values derived in cataclysmic variables are lower,
0.1--0.4 \citep*{King:2007a}.  For $\alpha=0.3$, the thermal
time-scale is about 60~min at the typical optical emission radius.
The viscous time-scale is much larger again, and too large to
associate with the lags we measure. Even for a very large scale height
of $H/R\sim0.2$, where $H$ is the disc half-thickness and $R$ is the
radius, we expect a viscous time-scale of around a day. For more
typically assumed $H/R\sim0.05$, the viscous time-scale becomes tens
of days (comparable to the decay time-scales of outbursts of transient
LMXBs; \citealt{King:1998a}).

\citet{Church:2012a} have argued that the Comptonization region
responsible for the hard X-rays is not a compact inner region, but
rather an extended thin, hot layer above the disc.  This is supported
by observations of progressive covering of the hard X-rays during
X-ray dips in dipping systems, permitting an estimate of the spatial
extent of the hard X-ray region.  Typical estimates for the radius for
the extended region are $1\times10^9$--$7\times10^{10}$~cm.  This is
at largest a factor of ten smaller than our estimate of the radius of
the region dominating optical emission, and could be much smaller than
this, so even in the \citet{Church:2012a} model we expect the
Comptonization region to be much more compact than the optical
emission region, and physically distinct; it does not overlay the
optical disc, for example, so in all models we expect the optical disc
emission to originate from a distinct region to the X-rays.

We have shown that the optical flux generally increases as the Z
diagram is traced from hard apex to soft apex and up the FB.  It has
often been suggested that the optical flux may be a better tracer of
mass accretion rate in Z sources than the X-ray flux is
\citep[e.g.][]{McNamara:2003a,OBrien:2004a}.  If this is the case,
then these observations vindicate the models where the Z diagram,
traced from HB to NB to FB is a monotonically increasing sequence in
accretion rate \citep[e.g.][]{Psaltis:1995a}.  Drawing this conclusion
does, however, rest on associating the optical flux with the mass
accretion rate.  It is far from clear that this is the case when
dramatic optical changes occur on time-scales much shorter than the
disc viscous time-scale.  Much of the optical emission in LMXBs, in
general, is attributed to reprocessing of X-ray emission
\citep{vanParadijs:1994a} and higher time resolution studies support
this \citep{MunozDarias:2007a}.  We certainly see evidence for this in
the variability in the FB, where good correlations are seen between
X-ray and optical. Moving to the NB, however, we find no positive
correlation.  It is then possible that the optically emitting regions
might be partially shielded from X-ray irradiation, which could result
in a lower overall optical flux, even at the same or higher mass
accretion rate. It is then possible that the optical flux is a poor
tracer of accretion rate. This would be consistent with several more
recent X-ray studies such as \citet*{Church:2006a} and
\citet{Church:2012a}, which locate the point of highest accretion rate
at the hard apex where X-ray flux is
maximized. \citet{Titarchuk:2014a} also suggest that the X-ray flux
should trace the accretion rate well. We will examine this possibility
further in the following section.

While the peak of the CCF in the high state is marginally consistent
with zero, it does suggest lags that are larger than expected from
purely light travel times in the binary ($\sim10$~s).  If the optical
emission does arise in X-ray reprocessing there can be an additional
delay due to the finite reprocessing time, dominated by the diffusion
time for optical photons to reach the surface after being deposited at
optical depth greater than unity.  This was considered in detail by
\citet*{Cominsky:1987a} who found that most of the response to X-ray
irradiation should be prompt, emerging within a second, but that for
reprocessing by harder photons the reprocessing time could increase
substantially, with some emission continuing to a few minutes after
the irradiation. This may contribute to the modest lags we see when
cross-correlating with hard X-ray light curves. Strong irradiation
could also lead to changes in the vertical structure of the outer disc
on the thermal time-scale. \citet{Scaringi:2015a} invoked a thermal
time-scale response to explain the $\sim12$~hr lag seen with respect
to MAXI data (see discussion in Section~\ref{CCFSection}, although
this requires lower viscosity parameters, $\alpha$, than we assumed
above. Given the estimates above of a thermal time-scale of about an
hour, the more natural feature to associate with the disc thermal
response might be the transition time-scales, with a full transition
between optical states taking from one to a few thermal time-scales.

The anti-correlation in the low-state is more challenging to explain.
\citet{McNamara:2003a} and \citet{Scaringi:2015a} attribute the
optical variability in the NB to the central region, and argue that
the optical depth becomes high enough to reprocess X-ray photons to
optical ones.  An increase in optical depth then causes a decrease in
X-rays and an increase in optical, i.e. an anti-correlation, as
observed.This interpretation is also consistent with the
anti-correlation being strongest at higher energies.  We will discuss
an alternative explanation for this in the following section.

\section{An Irradiative Explanation for the X-ray/Optical Connections in Sco~X-1}
\label{ModelSection}

We finish this work by attempting to tie together the various themes
that have been discussed into a coherent picture of the relationship
between X-ray and optical emission in Sco~X-1. There is a long history
of belief that the optical emission in luminous LMXBs such as Sco~X-1
is dominated by the outer accretion disc and secondary star, and more
specifically by thermal reprocessing of X-ray emission
\citep{vanParadijs:1994a}.  In this section, we will explore the
extent to which our observations, and in particular the bimodal
behaviour and anti-correlations, can still be explained within the thermal
reprocessing paradigm.

We begin by returning to the {\it Kepler} light curve shown in
Figure~\ref{LongLCFig}.  This, together with the flux histogram
(Figure~\ref{HistFig}) demonstrates a clear bimodal behaviour.  The
high state consists of a roughly flat plateau with fast flares and
slower dips.  The low state shows flat or U-shaped minima with slow
brightenings.  The low state brightenings and high state dips both
share a common time-scale, which is also the time-scale of full state
transitions. It is notably longer than the time-scales of high state
flares which appear to be a distinct phenomenon.  The high state
flares involve substantial increases in X-ray flux and more modest
correlated increases in optical flux. This behavior is consistent with
thermal reprocessing. As the system transitions to the low state, we
see a substantial decrease in optical flux at nearly constant X-ray
flux in all the energy bands studied. Quantitatively, the high state
peak in the optical histogram is at a 60~percent higher optical flux
than the low state. In contrast, depending on the energy band studied,
the average X-ray flux for observations during the high state in
Figure~\ref{IntensityFig} is only 10--22~percent higher than in the
low state. Furthermore, as Sco~X-1 moves from the soft apex up the
normal branch, the optical flux continues to decrease even while the
X-ray flux is {\em increasing}; this is the X-ray to optical
anti-correlation. Detailed spectral modeling by \citet{Church:2012a}
suggests that movement away from the soft apex along the FB and the NB
both involve increasing X-ray luminosity, and this conclusion is
supported by \citet{Titarchuk:2014a}.  The decrease in optical flux
cannot then simply represent reduced X-ray luminosity, so if it arises
by irradiation this must reflect a decreased irradiation efficiency in
the low state \citep[c.f.][]{Dubus:1999a} as a consequence of changes
in either the X-ray spectral shape (and hence albedo to those X-rays)
and/or illumination geometry.

A key observation here is that most of the change in optical flux
during a transition (complete or failed) between the high and low
states occurs at nearly constant X-ray flux
(Figure~\ref{IntensityFig}). X-ray data during times of optical
transition are also clearly located at the soft apex of the Z diagram,
and do not significantly extend on to either the flaring or normal
branch (Figure~\ref{ZFig}). The transitions, therefore, do not
correspond to significant changes in either the X-ray intensity or the
X-ray spectrum, so cannot be explained by either luminosity or
spectral shape changes. Even if spectral changes did come into play,
we have shown in Section~\ref{XraySpecSection} that the X-ray spectral
changes between high and low state are quite modest, at least from
2--40~keV, consistent with findings of \citet{Church:2012a} and
\citet{Titarchuk:2014a}. It thus seems that we must look to changes in
the illumination geometry to explain the optical state changes.

The overall light curve is quite suggestive of variable obscuration,
with the high state and low state corresponding to minimum and maximum
obscuration.  The dynamic range of the material causing the obscuration
may extend beyond that necessary to fully expose or fully obscure the
illuminating source, leading to saturated flat maxima in the high
state and, to a lesser extent, flat minima in the low state.  In the
clear state a partial obscuration can occur causing a dip, and in the
obscured state the obscuration can partially recede creating a
brightening, which is actually a reduction in the obscuration.  If the
Sco~X-1 behavior arises in a similar way then it is natural that the
time-scales for high state dip and low state brightening are similar,
as both arise on time-scales for changes in the obscuration.  In the
case of Sco~X-1 we are not suggesting that the optical light source
itself is being obscured, rather that obscuration in the inner disc is
blocking irradiation of the outer disc and possibly the companion. We
also emphasise that we are not suggesting that the more efficient
illumination in the high state causes flaring behaviour, rather that
when the inner accretion flow enters its high state configuration, two
effects are seen: it becomes unstable to flaring, and it illuminates
the outer disc more efficiently.

Such obscuration does arise in models of the inner disc of
Sco~X-1. For example, \citet{Titarchuk:2014a} identifies the dominant
Comptonization region as a vertically extended transition region
between the accretion disc and the neutron star. Much of the X-ray
flux that we see is scattered from the inner face of this region and
would not efficiently irradiate the outer disc. Equally, the extended
Comptonization region above the inner disc of \citet{Church:2012a}
would tend to scatter X-rays originating from near the neutron star
and reduce direct illumination of the outer disc.
\citet{Bradshaw:2003a} also invoke geometric changes in the accretion
configuration between states. A variety of behaviours could produce
the observed signatures. If there is a thick torus in the inner disc,
for example the transition region of \citet{Titarchuk:2014a}, then
decreasing the geometrical thickness of the torus would increase disc
irradiation. If there is clumpy absorbing material above the disc,
then decreasing the covering factor would increase irradiation of both
disc and companion. If the optical depth of scattering material above
the disc decreases then more direct irradiation would escape to
illuminate the outer disc and companion.  In any of these cases the
disc could only be irradiated by a small fraction of the X-ray flux
that emerges along our line of sight with the potential for large
variations in the illumination as the inner disc geometry changes. Of
the three examples, a torus of changing thickness would be expected to
affect the disc more than the companion, leading to changes in the
amplitude of the orbital modulation between states, inconsistent with
observations.  

It is at first puzzling that such structural changes should occur
suddenly at the soft apex, rather than while ascending either the
normal or flaring branch, but there is evidence for such abrupt
changes in the timing behavior of Sco~X-1. It is seen that dramatic
changes in the quasi-periodic oscillation frequencies can occur with
negligible movement in the Z diagram \citep{Dieters:2000a}, and that
the bottom NB and bottom FB have nearly identical energy spectra but
quite different time variability properties
\citep{Titarchuk:2014a}. As a result distinctly different normal and
flaring branch states can coexist at the soft apex for virtually the
same X-ray intensity and colours. Changes in timing behaviour should
reflect structural changes in the inner disc, and so could be
associated with the changes in illumination geometry that we infer.
The time-scales for NB--FB transitions can range from minutes for the
fastest timing excursions \citep{Dieters:2000a} up to more typical
transitions lasting tens of minutes or longer
\citep{Priedhorsky:1986a}. These are comparable to the shortest
transition time-scales that we see in the optical, but shorter than is
typical. The time taken to complete a full state transition,
$\sim1.7$~hours rising and $\sim2.0$~hours falling are quite similar
to our estimates of the disc thermal time-scale, so it may be that the
optical transitions also involve the disc adjusting to the change in
irradiation on the thermal time-scale.

Finally we note that extended periods of predominantly high state
behaviour last for about a week, quite comparable to the viscous
time-scale estimated above. If enhanced irradiation is responsible for
the changes in the high state, and leads to adjustments in the outer
disc vertical structure on the thermal time-scale, we would then
expect this to modulate the accretion flow through the disc. Such an
accretion rate change would take approximately the viscous time to
propagate to the inner disc where it could then end the predominantly
high state episode.

\section{Conclusions}
\label{ConclusionSection}

We have analyzed {\it Kepler} data from the K2 mission for Sco~X-1 in
conjunction with simultaneous {\it Fermi} GBM and MAXI X-ray data. The
overall light curve shows a bimodal form, with high optical fluxes
when the source is on the X-ray FB and lower fluxes on the NB. During
the optical high state we see a mix of fast flares and slower
dips. During the low state we see slower brightenings on similar
time-scales to the high state dips. The overall impression given is
that the brightenings and dips represent failed transitions between
the low and high state, whereas the faster flares are distinct.  We
correlate the {\it Kepler} light curves against simultaneous {\it
  Fermi} GBM and MAXI measurements of the hard X-ray flux.  We find a
clear relation between hard X-ray and optical flux, with the optical
positively correlated with hard X-rays in the high state and
anti-correlated in the low state. This can be seen both from flux-flux
diagrams, and from cross-correlation functions. We emphasize that the
cross-correlation behaviour is completely different in high and low
states, and so an average cross-correlation function combining states
is confused and should be interpreted with caution.

There is a clear orbital modulation, consistent with earlier
observations but of unprecedented quality. No compelling deviations
from an average sinusoidal form are found, and neither are significant
differences in the fractional amplitude of the orbital modulation
between high and low state observations.
 
We have analysed the statistical distribution of several
characteristic events in the light-curve for which {\em Kepler}
provides an unprecedented and uniform sample. Extended periods of
predominantly high state behaviour last for about a week, possibly
being shut off by changes on the disc viscous time-scale. In more
detail, uninterrupted high state periods have a log-normal
distribution with median duration of 4.7~hours. There appears to be a
characteristic time-scale for transitions between low and high states,
with these often showing an exponential rise and decay with
median times for full transitions of 1.7~hours when increasing flux
and 2.0~hours when decreasing. The slower decay time appears
statistically significant. These time-scales are common not only to
full transitions, but also to partial brightenings from the low state
and dips from the high state.  Finally, the high state flares show a
striking uniformity of amplitude with an apparent cut-off at a maximum
amplitude. The flares show no systematic asymmetry in time, and
durations (typically 5--20~min) that correlate positively with the
amplitude. They are likely to be produced by thermal reprocessing of
the X-ray flares for which the FB is named. High state
cross-correlation functions with GBM and MAXI show a sharp positive
response with delay of no more than a few minutes and marginally
consistent with zero. This is consistent with the thermal reprocessing
picture. A delay of a few minutes could arise from a finite
reprocessing time as hard X-ray photons deposit energy at significant
optical depth in the disc and companion star.

In the low state the optical anti-correlates with X-rays, with large
optical changes occurring during the transition for modest X-ray
changes. Both high state dips and low state brightenings seem to be
manifestations of failed state transitions. Deeper in the low state we
see larger X-ray changes for small optical ones, but the
anti-correlation is maintained.  Cross-correlation functions in the low
state show a pronounced anti-correlation feature, most prominent with
respect to GBM and to the MAXI highest energy band. The
anti-correlation is much broader than the high state correlation and
suggests a 30~min optical lag, although remains marginally consistent
with zero lag.

The clear separation of states using optical data can be exploited to
classify multi-wavelength data by state when other diagnostics
(e.g. X-ray colour-colour diagrams, or fast X-ray timing) are not
available. We demonstrate this by classifying individual GBM
occultation measurements into high, low and transition states and
constructing a hard X-ray spectrum for each. This is the first time
this has been possible with GBM data, as they provide no sensitivity
to soft X-ray colours or high frequency variability.

We suggest that the bimodal optical behaviour is caused by changes in
the efficiency with which the outer disc is irradiated with efficient
irradiation in the high state and reduced irradiation in the low
state. These changes could arise from increases in the thickness or
covering factor of obscuring material, or increases in scattering
optical depth which scatter material out of line of sight to the outer
disc.  The time-scales of optical transitions are comparable to, or
longer than the time-scales of transitions in X-ray timing behavior at
the soft apex of the Z diagram. Changes in X-ray timing signatures
presumably reflect changes in the accretion configuration near the
neutron star, and so can be expected to affect irradiation efficiency.
This then leads to dramatic changes in optical brightness when Sco~X-1
transitions between the different configurations of the normal and
flaring branch possibly also involving a thermal time-scale adjustment
in the outer disc vertical structure which smears out the transitions.

\section*{Acknowledgements}

This paper includes data collected by the {\it Kepler} and {\it Fermi}
missions. Funding for these missions is provided by the NASA Science
Mission directorate.  The {\it Kepler} data presented in this paper
were obtained from the Mikulski Archive for Space Telescopes
(MAST). STScI is operated by the Association of Universities for
Research in Astronomy, Inc., under NASA contract NAS5-26555. Support
for MAST for non-HST data is provided by the NASA Office of Space
Science via grant NNX13AC07G and by other grants and contracts.

This work made use of {\sc PyKE} \citep{Still:2012a}, a software
package for the reduction and analysis of {\it Kepler} data. This open
source software project is developed and distributed by the NASA
Kepler Guest Observer Office.

This research has made use of MAXI data provided by RIKEN, JAXA and
the MAXI team.  We are grateful to Tatehiro Mihara for providing us
with scan time data.

S.~S. acknowledges funding from the Alexander von Humboldt foundation.
We are grateful for helpful suggestions from our referee which
enriched the analysis in this paper.  This research has made use of
NASA's Astrophysics Data System.

\label{lastpage}

\end{document}